\documentclass[nonacm]{acmart}

\AtBeginDocument{%
  }

\usepackage{booktabs,tabularx}
\usepackage{booktabs}
\usepackage{placeins}
\usepackage{graphicx}
\usepackage{ragged2e}
\usepackage{pifont}
\usepackage{subfiles}
\usepackage{xcolor}
\usepackage{pgfplots}
\pgfplotsset{compat=1.18}
\usepgfplotslibrary{colorbrewer}
\usepackage{tikz}
\usetikzlibrary{external, arrows.meta, positioning, tikzmark, calc}
\pgfdeclarelayer{foreground}
\pgfsetlayers{main,foreground} 

\tikzset{
	graph/.style={draw, fill=remainingGColor, rounded corners=4.5mm, inner sep=2.5mm, align=center, minimum width=2.5cm},
	node/.style={circle, draw, fill=black, inner sep=0pt, minimum size=6pt},
	lnode/.style={node, label=left:#1},
	rnode/.style={node, label=right:#1},
	tnode/.style={node, label=above:#1},
	bnode/.style={node, label=below:#1},
	nodeR/.style={ node, draw=lightgray, fill=lightgray},
	nodeE/.style={ node, draw=redgray, fill=redgray, fill opacity=0.6, draw opacity=0.6},
	nodeI/.style={ node, draw=greengray, fill=greengray, fill opacity=0.6, draw opacity=0.6},
	weight/.style={node, rectangle, fill opacity=0, draw opacity=0, text opacity = 1,text=weightColor, yshift=0.4cm, font=\small},
	lweight/.style={node, rectangle, fill opacity=0, draw opacity=0, text opacity = 1,text=weightColor, xshift=-0.5cm, font=\small},
	rweight/.style={node, rectangle, fill opacity=0, draw opacity=0, text opacity = 1,text=weightColor, xshift=0.5cm, font=\small},
	weightR/.style={node, rectangle, fill opacity=0, draw opacity=0, text opacity = 1,text=redgray, yshift=0.5cm, font=\small},
	weightText/.style={node, rectangle, fill opacity=1, text opacity = 1,text=weightColor, fill=white, draw=white , font=\small},
	edge/.style={draw=black, fill=black, thick},
	edgeR/.style={ edge, draw=lightgray, fill=lightgray, dashed}
}

\newcommand{\remainingG}[2]{
	\begin{pgfonlayer}{foreground}
		\node[graph] at (#1) (box) {#2};
	\end{pgfonlayer}
}
\definecolor{remainingGColor}{RGB}{226,243,253}
\definecolor{weightColor}{RGB}{204,0,0}
\definecolor{darkgreen}{rgb}{0.0, 0.5, 0.0}
\definecolor{lightgreen}{rgb}{0.235, 0.7, 0.235}
\definecolor{green}{RGB}{0, 150, 130}
\definecolor{green70}{RGB}{76, 181, 167}
\definecolor{blue}{RGB}{70, 100, 170}
\definecolor{blue70}{RGB}{125,146,195}
\definecolor{blue50}{RGB}{162,177,212}                                                 \definecolor{blue30}{RGB}{199,208,229}
\definecolor{blue15}{RGB}{227,232,242}
\definecolor{lightgray}{rgb}{0.86,0.86,0.86}
\definecolor{greengray}{RGB}{193,228,224}
\definecolor{redgray}{RGB}{233,183,183}
\definecolor{red}{RGB}{204,0,0}
\definecolor{lipicsYellow}{RGB}{252,199,18}

\newcommand{\ie}{i.e.\ }
\newcommand{\etal}{et~al.}
\newcommand{\eg}{e.g.\ }
\newcommand{\I}{\mathcal{I}}
\newcommand{\tI}{\mathcal{\tilde{I}}}
\newcommand{\w}{\omega}
\newcommand{\aw}{\alpha_\omega}

\newcommand{\cliquer}{\textsc{Cliquer}}
\newcommand{\mwclq}{\textsc{MWCLQ}}
\newcommand{\wlmc}{\textsc{WLMC}}
\newcommand{\tsmmwc}{\textsc{TSM-MWC}}
\newcommand{\mwcredu}{\textsc{MWCRedu}}
\newcommand{\maxcliqueweight}{\textsc{MaxCliqueWeight}}

\newcommand{\mnts}{\textsc{MN/TS}}
\newcommand{\bls}{\textsc{BLS}}
\newcommand{\lsccbps}{\textsc{LSCC+BPS}}
\newcommand{\fastwclq}{\textsc{FastWClq}}
\newcommand{\rets}{\textsc{ReTS2}}
\newcommand{\rrwl}{\textsc{RRWL}}
\newcommand{\sccwalk}{\textsc{SCCWalk4L}}
\newcommand{\fastwclqn}{\textsc{FastWClq-V2}}
\newcommand{\mwcpeel}{\textsc{MWCPeel}}

\newcommand{\sbms}{\textsc{SBMS}}
\newcommand{\bmwvc}{\textsc{BMWVC}}

\newcommand{\aco}{\textsc{ACO}}
\newcommand{\acosee}{\textsc{ACO+SEE}}
\newcommand{\pbig}{\textsc{PBIG}}
\newcommand{\msits}{\textsc{MS-ITS}}
\newcommand{\dlswcc}{\textsc{DLSWCC}}
\newcommand{\fastwvc}{\textsc{FastWVC}}
\newcommand{\numwvc}{\textsc{NuMWVC}}
\newcommand{\dynwvc}{\textsc{DynWVC2}}
\newcommand{\maehts}{\textsc{MAE-HTS}}
\newcommand{\pgto}{\textsc{PGTO}}
\newcommand{\gnnvc}{\textsc{GNN-VC}}
\newcommand{\egmwvc}{\textsc{EG-MWVC}}

\newcommand{\kamis}{\textsc{KaMIS}}
\newcommand{\solve}{\textsc{Solve}}
\newcommand{\struction}{\textsc{Struction}}
\newcommand{\cbnr}{\textsc{C-B\&R}}
\newcommand{\lnr}{\textsc{LearnAndReduce}}

\newcommand{\plswis}{\textsc{PLS\_WIS}}
\newcommand{\ilsvnd}{\textsc{ILS-VND}}
\newcommand{\dttwo}{\textsc{DtTwo}}
\newcommand{\htwis}{\textsc{HtWIS}}
\newcommand{\metamis}{\textsc{METAMIS}}
\newcommand{\mmwis}{\textsc{$\textsc{m}^2$wis}}
\newcommand{\csearch}{\textsc{C-Search}}
\newcommand{\bsa}{\textsc{BSA}}
\newcommand{\hglv}{\textsc{HGLV}}
\newcommand{\chils}{\textsc{CHILS}}
\newcommand{\dynls}{\textsc{DynLS}}

\newtheoremstyle{custom}%
  {3pt}{3pt}
  {\itshape}
  {}
  {\bfseries}
  {}
  { }
  {\thmname{#1}\thmnumber{ #2}\thmnote{ \textnormal{(#3)}}}
\makeatother

\theoremstyle{custom}
\newtheorem{internalreduction}{Reduction}[subsection] 

\newenvironment{reduction}[1][\unskip]{
    \begin{internalreduction}[#1]
    \noindent\trivlist\item\vspace{-.5em}\ignorespaces
}{
    \endtrivlist
    \end{internalreduction}
}
\newenvironment{figreduction}[2][\unskip]{
\begin{internalreduction}[#1. Figure~\ref{fig:#2}]\phantomsection\label{red:#2}
    \noindent\trivlist\item\vspace{-.5em}\ignorespaces
}{
    \endtrivlist
    \end{internalreduction}
}

\newcommand{\reductiondetails}[3]{
\begin{tabular}{p{2.4cm}p{10.5cm}}
\\[-.5em]
    {Reduced Graph} & #1 \\
    {Offset} & #2 \\
    {Reconstruction} & #3 \\
\end{tabular}\\
}

\theoremstyle{acmdefinition} 
\newtheorem{remark}{Remark}[section]

\begin{document}

\title{A Comprehensive Survey of Data Reduction Rules for the Maximum Weighted Independent Set Problem}

\author{Ernestine Gro{\ss}mann}
\authornote{Corresponding author}
\email{e.grossmann@informatik.uni-heidelberg.de}
\orcid{0000-0002-9678-0253}
\author{Kenneth Langedal}
\orcid{0009-0001-6838-4640}
\email{kenneth.langedal@informatik.uni-heidelberg.de}
\author{Christian Schulz}
\orcid{0000-0002-2823-3506}
\email{christian.schulz@informatik.uni-heidelberg.de}
\affiliation{%
    \institution{Faculty of Mathematics and Computer Science, Heidelberg University}
    \country{Germany}
}

\renewcommand{\shortauthors}{E. Gro{\ss}mann, K. Langedal, C. Schulz}

\begin{abstract}
    The \textsc{Maximum Weight Independent Set} problem, as well as its related problems such as \textsc{Minimum Weight Vertex Cover}, are fundamental $\mathsf{NP}$-hard problems with numerous practical applications. Due to their computational complexity, a variety of data reduction rules have been proposed in recent years to simplify instances of these problems, enabling exact solvers and heuristics to handle them more effectively. Data reduction rules are polynomial time procedures that can reduce an instance while ensuring that an optimal solution on the reduced instance can be easily extended to an optimal solution for the original instance. Data reduction rules have proven to be especially useful in branch-and-reduce methods, where successful reductions often lead to problem instances that can be solved exactly. This survey provides a comprehensive overview of data reduction rules for the \textsc{Maximum Weight Independent Set} problem. We also provide a reference implementation for these reductions. This survey will be updated as new reduction techniques are developed, serving as a centralized resource for researchers and practitioners.
    
    \vspace{1em}
    \noindent\textbf{Source Code:} \url{https://github.com/KarlsruheMIS/DataReductions}
\end{abstract}


\maketitle

\section{Introduction}
    
This survey presents a comprehensive overview of exact data reduction rules for the \textsc{Maximum Weight Independent Set} problem in practice, along with a reference implementation. Data reduction rules are polynomial time procedures that can reduce the size of a given input instance. After applying exact data reductions, an optimal solution on the reduced instance can be easily reconstructed to an optimal solution on the original graph.

An \textit{independent set} (IS) for a given graph $G(V,\,E)$ is defined as a subset ${\I\subseteq V}$ of vertices such that each pair of vertices in $\I$ are non-adjacent. In the \textsc{Maximum Independent Set} (MIS) problem, the task is to find an IS with the highest possible cardinality. A closely related problem to MIS is the \textsc{Minimum Vertex Cover} (MVC) problem. A \textit{vertex cover} $C \subseteq V$ is a set of vertices that cover all the edges, where an edge is \textit{covered} if it is incident to one vertex in the set~$C$. Note that for a maximum independent set $\I$ of $G$, $V \setminus \I$ is a minimum vertex cover, making MIS and MVC complementary problems. By transforming the graph~$G$ into the complement graph $\overline{G}$, the MIS problem in $G$ becomes the \textsc{Maximum Clique} problem in $\overline{G}$. A \textit{clique} is a subset of pairwise adjacent vertices, and the \textsc{Maximum Clique} (MC) problem is to find a clique of maximum cardinality. Despite MIS and MC also being complementary problems, using an MC algorithm to solve the MIS problem on a sparse graph $G$ is impractical since the complement $\overline{G}$ can be very dense and, therefore, unlikely to fit in memory for all but the smallest instances.

For a weighted graph ${G=(V, E, \omega)}$ with positive vertex weights given by a function $\omega:V \rightarrow \mathbb{R}^{+}$, the \textsc{Maximum Weight Independent Set} problem asks for an independent set $\I$ with maximum weight $\omega(\I) = \sum_{v \in \I} \omega(v)$. As with the unweighted problems, the weighted versions \textsc{Maximum Weight Independent Set} (MWIS), \textsc{Minimum Weight Vertex Cover} (MWVC), and \textsc{Maximum Weight Clique} (MWC) are also complementary. For example, if we find an MWIS $\I$ of $G$, we have simultaneously found an MWVC of $G$ by taking the vertices $V \setminus \I$, and an MWC in $\overline{G}$ using the same vertices $\I$. The MWIS problem, as well as the related problems addressed above, have several applications such as long-haul vehicle routing \cite{metamis}, the winner determination problem~\cite{wu2015solving}, or prediction of structural and functional \hbox{sites in proteins~\cite{mascia2010predicting}.}

Since MWIS, as well as its related problems, are $\mathsf{NP}$-hard~\cite{garey1974}, several new data reduction rules have been presented in recent years. These data reduction rules are polynomial time procedures that can reduce the size of an instance while ensuring that an optimal solution to the reduced instance can be easily extended to an optimal solution for the original instance. The notion of data reduction rules is often used in theoretical, fixed-parameter tractable algorithms (FPT). From a theoretical perspective, these algorithms can solve $\mathsf{NP}$-hard problems efficiently if some problem parameter $k$ is small. A typical problem parameter is the solution size. For FPT algorithms, reduction rules are utilized in so-called \emph{kernelization} routines, which reduce the input instance in polynomial time to a \emph{kernel}. A kernel is equivalent to the original input instance in that an optimal solution on the kernel can be extended to an optimal solution on the original instance. Furthermore, the size of the kernel is bounded by a computable function $f$, which is only dependent on the parameter $k$. For example, for the MVC problem, we can reduce an instance to a kernel with at most $2 \cdot k$ vertices~\cite{hartmanis2006texts}, which makes MVC an FPT problem. However, the MIS problem is most likely not FPT~\cite{downey1995fixed}. Unlike the theoretical perspective, MVC and MIS are considered equivalent in practice because by solving one, we implicitly solve the other.

Data reductions can be useful in practice even if the size of the reduced graph can not be bounded by any computable function $f$ depending on a parameter $k$. Therefore, after the first reduction rules developed for FPT algorithms, practical reductions have recently been developed without focusing on theoretical guarantees. These practical reductions have led to significant improvements in the performance of various exact algorithms and heuristics.
    
Exact solvers such as \textit{branch-and-reduce} methods \cite{gellner2021boosting, hespe2020wegotyoucovered, lamm2019exactly} often solve medium-sized instances in practice using reduction rules combined with branch-and-bound to further reduce the graph between branches. For branch-and-reduce solvers, it has been observed that if data reductions work well, then the instance is likely to be solved. If data reductions do not work that well, \ie the size of the reduced graph is similar to the original graph, then the instance can often not be solved. Despite recent improvements, such as the struction algorithm by~\cite{gellner2021boosting} that manages to solve several large instances to optimality, many publicly available instances remain unsolved.

Data reduction rules also play an important role in many heuristics, such as reduce-and-peel approaches as shown \hbox{in \cite{gu2021towards} or} \cite{DBLP:conf/icde/ZhengGPY20}. In the \textit{reduce-and-peel} approach, the graph is reduced all the way to zero vertices while using a heuristic tie-breaking mechanism to ensure continuous progress. This results in a heuristic where reasonably good solutions can be computed quickly. However, they can also be applied in more sophisticated approaches, as shown in \cite{mmwis, langedal2022efficient, li2020numwvc}.

Given the numerous new data reduction rules recently proposed for these problems, we aim to collect and present them all in a single paper, which will be updated as new reduction rules are introduced. Furthermore, we present an overview and brief description of all solvers for the MWC, MWVC, and MWIS problems. To show how the state-of-the-art solvers evolved over time, we include a high-level visualization that illustrates which solvers were used in the experimental evaluation of new solvers.

\section{Preliminaries}
In this work, a graph $G=(V,E)$ is an undirected graph with
${n=|V|}$ and ${m = |E|}$, where ${V =\{0,...,n-1\}}$. The neighborhood $N(v)$ of a vertex $v \in V$ is defined as $N(v) = \{u \in V \mid \{u,v\} \in E\}$.
Additionally, $N[v]=N(v) \cup \{v\}$. The same sets are defined for the neighborhood $N(U)$ of a set of vertices ${U \subseteq V}$, \ie ${N(U) = \cup_{v \in U} N(v)\setminus U}$
and $N[U] = N(U) \cup U$. The degree of a vertex $\mathrm{deg}(v)$ is defined as the number of its neighbors $\mathrm{deg}(v)=|N(v)|$. The complement graph is defined as ${\overline{G}=(V,\overline{E})}$, where ${\overline{E}=\{\{u,v\} \mid \{u,v\} \notin E\}}$ is the set of edges not present in $G$.
A set $\I \subseteq V$ is called \textit{independent set} (IS) if for all vertices $v,u \in \I$ there is no edge $\{v,u\} \in E$. For a given IS~$\I$ a vertex $v \notin \I$ is called \textit{free}, if $\I \cup \{v\}$ is still an independent set. If a vertex $v \notin \I$ is not free, the number of neighbors it has in $\I$ is called its \textit{tightness}. An IS is called \textit{maximal} if there are no free vertices.
The \textsc{Maximum Independent Set} (MIS) problem is that of finding an IS with maximum cardinality.
The \textsc{Maximum Weight Independent Set} (MWIS) problem is that of finding an IS with maximum weight. The weight of an independent set $\I$ is defined as $\omega(\I) = \sum_{v \in \I}\omega(v)$ and $\alpha_\omega(G)$ denotes the weight of an MWIS of $G$. The complement of an independent set is a \textit{vertex cover}, \ie a subset ${C \subseteq V}$ such that $\forall \{u ,v\} \in E \implies u \in C \vee v \in C$. In other words, every edge $e \in E$ is covered by at least one vertex $v \in C$, where an edge is \textit{covered} if it is incident to one vertex in the set~$C$. The \textsc{Minimum Vertex Cover} problem, defined as looking for a vertex cover with minimum cardinality, is thereby complementary to the MIS problem. Another closely related concept are cliques. A \textit{clique} is a set $Q \subseteq V$ such that all vertices are pairwise adjacent. A clique in the complement graph $\overline{G}$ corresponds to an independent set in the original graph~$G$. A vertex is called \textit{simplicial}, when its neighborhood forms a clique.

\section{Related Work}\label{sec:related_work}

    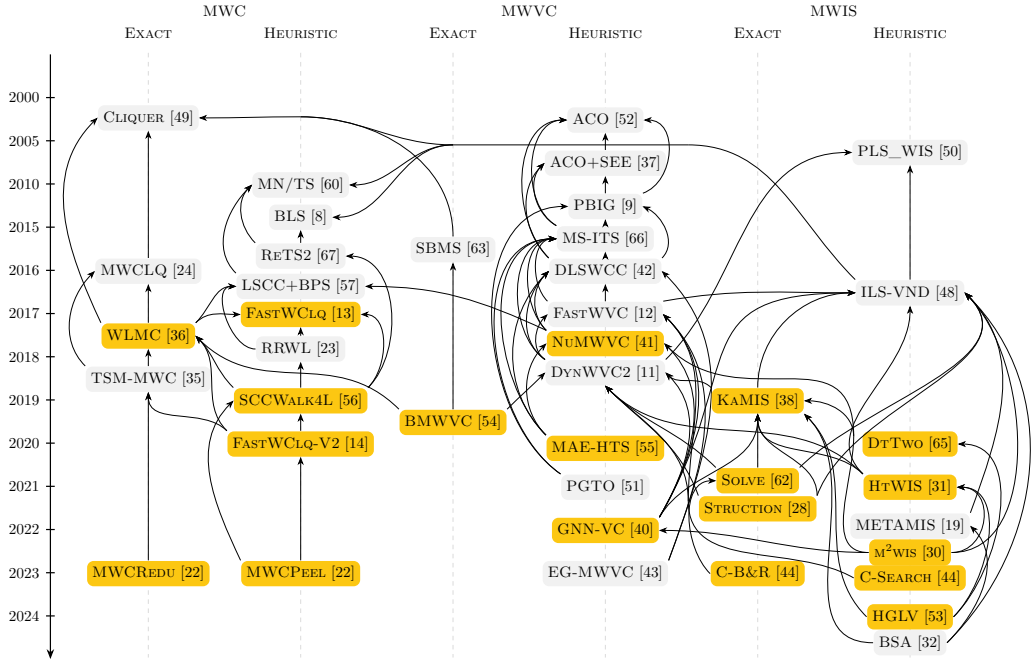
\begin{figure}[ht]
        \centering
        \resizebox{\textwidth}{!}{
        \begin{tikzpicture}[yscale = 1.5,
                red/.style={fill=lipicsYellow, thick},
                every node/.style={anchor=center}]
            \begin{scope}[xshift=-1.cm]
                \path[-{Stealth[scale=1.5]}, thick] (-3.5,-1) edge (-3.5, -16.);

                \foreach[evaluate={\l = int(5*\x + 2000)}] \x in {0,...,3} {
                        \path[-, thick] (-3.6, -\x - 2) edge (-3.4, -\x - 2);
                        \node[label=left:\large\l] (t_\x) at (-3.5, -\x -2) {};
                    }
                \foreach[evaluate={\l = int(\x + 2016)}] \x in {0,...,8} {
                        \path[-, thick] (-3.6, -\x - 6) edge (-3.4, -\x - 6);
                        \node[label=left:\large\l] (t_\x) at (-3.5, -\x -6) {};
                    }
            \path[-, thick] (-3.6, - 15.5) edge (-3.4, -15.5);
            \node[label=left:\large 2025] (t_2025) at (-3.5, -15.5) {};
            \end{scope}
            \node[label=center:\Large\textsc{MWC}] (mwc) at (0, 0) {};
            \node[label=center:\large\textsc{Exact}, below left = 0.75 and 1.95 of mwc.center, inner sep=13, anchor=center] (mwc-e) {};
            \node[label=center:\large\textsc{Heuristic}, below right = 0.75 and 1.95 of mwc.center, inner sep=13, anchor=center] (mwc-h) {};
            \path[-, dashed, lightgray!70!black] (mwc-e.south) edge ++(0, -15);
            \path[-, dashed, lightgray!70!black] (mwc-h.south) edge ++(0, -15);

            \node[label=center:\Large\textsc{MWVC}] (mwvc) at (8, 0) {};
            \node[label=center:\large\textsc{Exact}, below left = 0.75 and 1.95 of mwvc.center, inner sep=13, anchor=center] (mwvc-e) {};
            \node[label=center:\large\textsc{Heuristic}, below right = 0.75 and 1.95 of mwvc.center, inner sep=13, anchor=center] (mwvc-h) {};
            \path[-, dashed, lightgray!70!black] (mwvc-e.south) edge ++(0, -15);
            \path[-, dashed, lightgray!70!black] (mwvc-h.south) edge ++(0, -15);

            \node[label=center:\Large\textsc{MWIS}] (mwis) at (16, 0) {};
            \node[label=center:\large\textsc{Exact}, below left = 0.75 and 1.95 of mwis.center, inner sep=13, anchor=center] (mwis-e) {};
            \node[label=center:\large\textsc{Heuristic}, below right = 0.75 and 1.95 of mwis.center, inner sep=13, anchor=center] (mwis-h) {};
            \path[-, dashed, lightgray!70!black] (mwis-e.south) edge ++(0, -15);
            \path[-, dashed, lightgray!70!black] (mwis-h.south) edge ++(0, -15);

            \begin{scope}[every node/.style={rectangle, align=center, draw=none, thick, fill=lightgray!70, rounded corners=1.5mm}]

                \node[below = 1.8 of mwc-e] (cliquer) {\cliquer~\cite{ostergaard1999new}}; 
                \node[fill=none,left = 0.8 of cliquer] (cliquer1) {}; 
                \node[below = 7.125 of mwc-e] (mwclq) {\mwclq~\cite{fang2016exact}}; 
                \node[red, below = 9.375 of mwc-e] (wlmc) {\wlmc~\cite{jiang2017exact}}; 
                \node[below = 10.875 of mwc-e] (tsmmwc) {\tsmmwc~\cite{jiang2018two}}; 
                \node[red, below = 17.625 of mwc-e] (mwcredu) {\mwcredu~\cite{erhardt2023improved}}; 
                \node[below = 19.125 of mwc-e] (maxcliqueweight) {\maxcliqueweight~\cite{rozman2024exact}}; 

                \node[below = 4.125 of mwc-h] (mnts) {\mnts~\cite{wu2012multi}}; 
                \node[below = 5.25 of mwc-h] (bls) {\bls~\cite{benlic2013breakout}}; 
                \node[below = 6.6 of mwc-h] (rets) {\rets~\cite{zhou2017push}}; 
                \node[below = 7.65 of mwc-h] (lsccbps) {\lsccbps~\cite{wang2016two}}; 
                \node[red, below = 8.625 of mwc-h] (fastwclq) {\fastwclq~\cite{cai2016fast}}; 
                \node[below = 9.825 of mwc-h] (rrwl) {\rrwl~\cite{fan2017restart}}; 
                \node[red, below = 11.625 of mwc-h] (sccwalk) {\sccwalk~\cite{wang2020sccwalk}}; 
                \node[red, below = 13.125 of mwc-h] (fastwclqn) {\fastwclqn~\cite{cai2021semi}}; 
                \node[red, below = 17.625 of mwc-h] (mwcpeel) {\mwcpeel~\cite{erhardt2023improved}}; 

                \path[-{Stealth[scale=1.5]}] (mwclq) edge[out=90, in=270] (cliquer);

                \path[-{Stealth[scale=1.5]}] (rets.north west) edge[out=135, in=225] (mnts.west);
                \path[-{Stealth[scale=1.5]}] (rets.north) edge[out=90, in=270] (bls.south);

                \path[-{Stealth[scale=1.5]}] (lsccbps.north west) edge[out=135, in=225] (mnts.west);

                \path[-{Stealth[scale=1.5]}] (wlmc.north west) edge[out=115, in=225] (cliquer.west);
                \path[-{Stealth[scale=1.5]}] (wlmc) edge[out=90, in=270] (mwclq);
                \path[-{Stealth[scale=1.5]}] (wlmc.north east) edge[out=45, in=180] (lsccbps.west);
                \path[-{Stealth[scale=1.5]}] (wlmc.north east) edge[out=45, in=180] (fastwclq);

                \path[-{Stealth[scale=1.5]}] (rrwl.west) edge[out=180, in=180] (lsccbps.west);
                \path[-{Stealth[scale=1.5]}] (rrwl) edge[out=90, in=270] (fastwclq);

                \path[-{Stealth[scale=1.5]}] (sccwalk.north west) edge[out=135, in=-45] (wlmc.east);
                \path[-{Stealth[scale=1.5]}] (sccwalk) edge[out=90, in=270] (rrwl);
                \path[-{Stealth[scale=1.5]}] (sccwalk.north east) edge[out=45, in=0] (fastwclq);
                \path[-{Stealth[scale=1.5]}] (sccwalk.north east) edge[out=45, in=0] (rets);

                \path[-{Stealth[scale=1.5]}] (tsmmwc.north west) edge[out=135, in=225] (mwclq.west);
                \path[-{Stealth[scale=1.5]}] (tsmmwc.north) edge[out=90, in=270] (wlmc.south);

                \path[-{Stealth[scale=1.5]}] (fastwclqn.north west) edge[out=115, in=-45] (wlmc.east);
                \path[-{Stealth[scale=1.5]}] (fastwclqn.north west) edge[out=135, in=270] (tsmmwc);
                \path[-{Stealth[scale=1.5]}] (fastwclqn) edge[out=90, in=270] (sccwalk);

                \path[-{Stealth[scale=1.5]}] (mwcredu) edge[out=90, in=270] (tsmmwc);
                \path[-{Stealth[scale=1.5]}] (mwcpeel.north west) edge[out=115, in=225] (sccwalk.west);
                \path[-{Stealth[scale=1.5]}] (mwcpeel) edge[out=90, in=270] (fastwclqn);

                \path[-] (maxcliqueweight.north west) edge[out=90, in=270] (cliquer1.center);
                \path[-{Stealth[scale=1.5]}] (cliquer1.center) edge[out=0, in=180] (cliquer.west);

                \node[below = 6.375 of mwvc-e] (sbms) {\sbms~\cite{xu2016new}}; 
                \node[below = 12.375 of mwvc-e, red] (bmwvc) {\bmwvc~\cite{wang2019exact}}; 

                \node[below = 1.875 of mwvc-h] (aco) {\aco~\cite{shyu2004ant}}; 
                \node[below = 3.375 of mwvc-h] (acosee) {\acosee~\cite{jovanovic2011ant}}; 
                \node[below = 4.875 of mwvc-h] (pbig) {\pbig~\cite{bouamama2012population}}; 
                \node[below = 6.00 of mwvc-h] (msits) {\msits~\cite{zhou2016multi}}; 
                \node[below = 7.125 of mwvc-h] (dlswcc) {\dlswcc~\cite{li2016efficient}}; 
                \node[below = 8.625 of mwvc-h] (fastwvc) {\fastwvc~\cite{cai2019towards}}; 
                \node[below = 9.6375 of mwvc-h, red] (numwvc) {\numwvc~\cite{li2020numwvc}}; 
                \node[below = 10.65 of mwvc-h] (dynwvc) {\dynwvc~\cite{cai2018improving}}; 
                \node[below = 13.275 of mwvc-h, red] (maehts) {\maehts~\cite{wang2021fast}}; 
                \node[below = 14.625 of mwvc-h] (pgto) {\pgto~\cite{qiu2022population}}; 
                \node[below = 16.125 of mwvc-h, red] (gnnvc) {\gnnvc~\cite{langedal2022efficient}}; 
                \node[below = 17.625 of mwvc-h] (egmwvc) {\egmwvc~\cite{li2023evolutionary}}; 

                \path[-{Stealth[scale=1.5]}] (acosee) edge[out=90, in=270] (aco);

                \path[-{Stealth[scale=1.5]}] (pbig.east) edge[out=20, in=0, looseness=1.2] (aco.east);
                \path[-{Stealth[scale=1.5]}] (pbig) edge[out=90, in=270] (acosee);

                \path[-{Stealth[scale=1.5]}] (msits.north west) edge[out=150, in=180] (aco.west);
                \path[-{Stealth[scale=1.5]}] (msits.north west) edge[out=150, in=225] (acosee.west);
                \path[-{Stealth[scale=1.5]}] (msits) edge[out=90, in=270] (pbig);

                \path[-{Stealth[scale=1.5]}] (dlswcc.north west) edge[out=135, in=180] (aco);
                \path[-{Stealth[scale=1.5]}] (dlswcc.north west) edge[out=135, in=225] (acosee.west);
                \path[-{Stealth[scale=1.5]}] (dlswcc.north east) edge[out=45, in=335] (pbig.east);
                \path[-{Stealth[scale=1.5]}] (dlswcc) edge[out=90, in=270] (msits);

                \path[-{Stealth[scale=1.5]}] (fastwvc.north west) edge[out=135, in=180] (msits.west);
                \path[-{Stealth[scale=1.5]}] (fastwvc) edge[out=90, in=270] (dlswcc);

                \path[-{Stealth[scale=1.5]}] (numwvc.west) edge[out=135, in=180] (msits.west);
                \path[-{Stealth[scale=1.5]}] (numwvc.west) edge[out=155, in=0] (lsccbps.east);
                \path[-{Stealth[scale=1.5]}] (numwvc.west) edge[out=135, in=225] (dlswcc.west);

                \path[-{Stealth[scale=1.5]}] (dynwvc.west) edge[out=145, in=180] (msits.west);
                \path[-{Stealth[scale=1.5]}] (dynwvc.west) edge[out=145, in=225] (dlswcc.west);
                \path[-{Stealth[scale=1.5]}] (dynwvc.west) edge[out=145, in=225] (fastwvc.west);

                \path[-{Stealth[scale=1.5]}] (bmwvc) edge[out=90, in=270] (sbms);
                \path[-{Stealth[scale=1.5]}] (bmwvc.north west) edge[out=135, in=-45, looseness=1.1] (wlmc.east);
                \path[-{Stealth[scale=1.5]}] (bmwvc.north east) edge[out=45, in=225] (dynwvc.west);

                \path[-{Stealth[scale=1.5]}] (maehts.north west) edge[out=135, in=180] (msits.west);
                \path[-{Stealth[scale=1.5]}] (maehts.north west) edge[out=135, in=225] (dlswcc.west);
                \path[-{Stealth[scale=1.5]}] (maehts.north west) edge[out=135, in=225] (numwvc.west);

                \path[-{Stealth[scale=1.5]}] (pgto.north west) edge[out=155, in=180] (msits.west);
                \path[-{Stealth[scale=1.5]}] (pgto.north west) edge[out=155, in=180] (pbig.west);

                \path[-{Stealth[scale=1.5]}] (gnnvc.north east) edge[out=60, in=-45] (fastwvc.east);
                \path[-{Stealth[scale=1.5]}] (gnnvc.north east) edge[out=60, in=-45] (numwvc.east);
                \path[-{Stealth[scale=1.5]}] (gnnvc.north east) edge[out=60, in=-45] (dynwvc.east);

                \path[-{Stealth[scale=1.5]}] (egmwvc.north east) edge[out=70, in=-45] (fastwvc.east);
                \path[-{Stealth[scale=1.5]}] (egmwvc.north east) edge[out=70, in=-45] (dlswcc.east);

                \node[below = 11.625 of mwis-e, red] (kamis) {\kamis~\cite{lamm2019exactly}}; 
                \node[below = 14.4 of mwis-e, red] (solve) {\solve~\cite{xiao2021efficient}}; 
                \node[below = 15.375 of mwis-e, red] (struction) {\struction~\cite{gellner2021boosting}}; 
                \node[below = 17.625 of mwis-e, red] (cbnr) {\cbnr~\cite{liu2023application}}; 

                \node[below = 1.5 of mwis-h] (plswis) {\plswis~\cite{pullan2009optimisation}}; 
                \node[below = 7.875 of mwis-h] (ilsvnd) {\ilsvnd~\cite{nogueira2018hybrid}}; 
                \node[below = 13.125 of mwis-h, red] (dttwo) {\dttwo~\cite{zheng2020efficient}}; 
                \node[below = 14.625 of mwis-h, red] (htwis) {\htwis~\cite{gu2021towards}}; 
                \node[below = 15.975 of mwis-h] (metamis) {\metamis~\cite{dong2022metaheuristic}}; 
                \node[below = 16.875 of mwis-h, red] (mmwis) {\mmwis~\cite{grossmann2023finding}}; 
                \node[below = 17.775 of mwis-h, red] (csearch) {\csearch~\cite{liu2023application}}; 
                \node[below = 19.125 of mwis-h, red] (hglv) {\hglv~\cite{tan2024efficient}}; 
                \node[below = 20.025 of mwis-h] (bsa) {\bsa~\cite{haller2024bregman}}; 
                \node[below = 21.075 of mwis-h] (chils) {\chils~\cite{grossmann2025concurrent}}; 
                \node[red,below = 21.75 of mwis-h] (dynls) {\dynls~\cite{zhu2025dynamic}}; 

                \path[-{Stealth[scale=1.5]}] (fastwvc.north east) edge[out=10, in=180] (ilsvnd.west);
                \path[-{Stealth[scale=1.5]}] (dynwvc.east) edge[out=45, in=180] (plswis.west);
                \path[-{Stealth[scale=1.5]}] (gnnvc.north east) edge[out=60, in=180, looseness=1.7] (ilsvnd.west);
                \path[-{Stealth[scale=1.5]}] (gnnvc.north east) edge[out=60, in=270,looseness=0.7] (kamis.south);

                \path[-{Stealth[scale=1.5]}] (ilsvnd.north) edge[out=90, in=270] (plswis.south);
                \node[below left = 3 and 1 of mwis-e, fill=none] (ilsvndt1) {};
                \node[below = 3 of mwvc-e, fill=none] (ilsvndt2) {};
                \node[below = 2.025 of mwc-h, fill=none] (ilsvndt3) {};
                \path[-] (ilsvnd.north west) edge[out=135, in=0] (ilsvndt1.center);
                \path[-] (ilsvndt1.center) edge[out=180, in=0] (ilsvndt2.center);
                \path[-] (ilsvndt2.center) edge[out=180, in=0] (ilsvndt3.center);
                \path[-] (sbms.north) edge[out=90, in=0] (ilsvndt3.center);
                \path[-{Stealth[scale=1.5]}] (ilsvndt3.center) edge[out=180, in=0] (cliquer.east);
                \path[-{Stealth[scale=1.5]}] (ilsvndt2.center) edge[out=180, in=0] (mnts.east);
                \path[-{Stealth[scale=1.5]}] (ilsvndt2.center) edge[out=180, in=0] (bls.east);

                \path[-{Stealth[scale=1.5]}] (kamis.west) edge[out=180, in=-45] (dynwvc.east);
                \path[-{Stealth[scale=1.5]}] (kamis.north) edge[out=90, in=180] (ilsvnd.west);

                \path[-{Stealth[scale=1.5]}] (htwis.north west) edge[out=135, in=270] (ilsvnd.south);
                \path[-{Stealth[scale=1.5]}] (htwis.north west) edge[out=135, in=-45] (dynwvc.south east);

                \path[-{Stealth[scale=1.5]}] (solve.north) edge[out=90, in=270] (kamis.south);
                \path[-{Stealth[scale=1.5]}] (solve.north east) edge[out=47, in=-45] (ilsvnd.east);
                \path[-{Stealth[scale=1.5]}] (solve.north west) edge[out=135, in=315] (dynwvc.south east);

                \path[-{Stealth[scale=1.5]}] (struction.north east) edge[out=90, in=270] (kamis.south);
                \path[-{Stealth[scale=1.5]}] (struction.north east) edge[out=79, in=-45] (ilsvnd.east);
                \path[-{Stealth[scale=1.5]}] (struction.north west) edge[out=110, in=315] (dynwvc.south east);

                \path[-{Stealth[scale=1.5]}] (metamis.north east) edge[out=70, in=-45] (ilsvnd.east);

                \node[below left = 14.7 and 5.5 of mwis-h, fill=none] (csearch1) {};
                \node[below left = 12.375 and 0.8 of mwis-h, fill=none] (mmwis1) {};
                \node[below left = 18. and 1.5 of mwis-h, fill=none] (kamis1) {};
                \node[below left = 18. and 1.5 of mwis-h, fill=none] (kamis1) {};
                \node[below left = 16. and 1. of mwis-e, fill=none] (solve1) {};
                \node[below left = 15. and 1.5 of mwis-h, fill=none] (kamis2) {};
                \node[below left = 12. and 1.5 of mwis-h, fill=none] (kamis3) {};

                \path[-] (mmwis.west) edge[out=180, in=270] (mmwis1.center);
                \path[-] (mmwis.west) edge[out=180, in=-80] (kamis2.center);
                \path[-{Stealth[scale=1.5]}] (mmwis.east) edge[out=0, in=0, looseness=1.5] (htwis.east);
                \path[-{Stealth[scale=1.5]}] (mmwis1.center) edge[out=90, in=-45, looseness=1.3] (numwvc.east);
                \path[-{Stealth[scale=1.5]}] (mmwis.east) edge[out=0, in=-45, looseness=0.9] (ilsvnd.east);
                \path[-{Stealth[scale=1.5]}] (mmwis.west) edge[out=180, in=0,looseness=0.5] (gnnvc.east);
                \path[-{Stealth[scale=1.5]}] (mmwis.west) edge[out=180, in=325] (struction.south east);

                \path[-{Stealth[scale=1.5]}] (solve1.center) edge[out=90, in=180] (solve.west);
                \path[-] (cbnr.west) edge[out=135, in=270] (solve1.center);
                \path[-] (htwis.north west) edge[out=135, in=-45] (kamis3.center);

                \path[-] (csearch.north west) edge[out=150, in=270] (csearch1.center);
                \path[-{Stealth[scale=1.5]}] (csearch1.center) edge[out=90, in=-45] (fastwvc.east);
                \path[-{Stealth[scale=1.5]}] (csearch1.center) edge[out=90, in=-45] (dynwvc.south east);

                \path[-{Stealth[scale=1.5]}] (kamis3.center) edge[out=125, in=0, looseness=0.8] (kamis.east);
                \path[-] (kamis1.center) edge[out=90, in=-90, looseness=0.5] (kamis2.center);
                \path[-] (kamis2.center) edge[out=90, in=-80, looseness=0.5] (kamis3.center);
                \path[-] (hglv.west) edge[out=180, in=-90] (kamis1.center);
                \path[-{Stealth[scale=1.5]}] (hglv.east) edge[out=45, in=0] (htwis.east);
                \path[-{Stealth[scale=1.5]}] (hglv.east) edge[out=45, in=0] (dttwo.east);

                \path[-] (bsa.west) edge[out=180, in=-90] (kamis1.center);
                \path[-{Stealth[scale=1.5]}] (bsa.east) edge[out=45, in=-45] (ilsvnd.east);
                \path[-{Stealth[scale=1.5]}] (bsa.east) edge[out=45, in=-45] (metamis.east);

                \path[-{Stealth[scale=1.5]}] (chils.east) edge[out=45, in=-45] (ilsvnd.east);
                \path[-{Stealth[scale=1.5]}] (chils.east) edge[out=45, in=-45] (metamis.east);
                \path[-{Stealth[scale=1.5]}] (chils.east) edge[out=45, in=0] (mmwis.east);
                \path[-{Stealth[scale=1.5]}] (chils.east) edge[out=45, in=0] (htwis.east);
                \path[-{Stealth[scale=1.5]}] (chils.north) edge[out=90, in=270] (bsa.south);

                \path[-{Stealth[scale=1.5]}] (dynls.east) edge[out=45, in=-45] (ilsvnd.east);
                \path[-{Stealth[scale=1.5]}] (dynls.east) edge[out=45, in=-45] (mmwis.east);
                \path[-{Stealth[scale=1.5]}] (dynls.east) edge[out=45, in=0] (htwis.east);
                \path[-{Stealth[scale=1.5]}] (dynls.west) edge[out=180, in=270] (struction.345);
                \path[-] (dynls.west) edge[out=180, in=270] (solve1.center);

            \end{scope}
        \end{tikzpicture}}
        \caption{This figure illustrates the history of MWC, MWVC, and MWIS solvers. The left axis gives a rough overview of publication years. A directed edge from a solver indicates a comparison made to another solver in the experimental evaluation. For example, the edge from \mwcredu{} to \tsmmwc{} indicates that \mwcredu{} used \tsmmwc{} in the experimental evaluation. The solvers that are highlighted in yellow are \noindent\colorbox{lipicsYellow}{using data reductions.}}
        \label{fig:history}
    \end{figure}

We give a brief overview of existing work on the \textsc{Maximum Weight Clique} (MWC), \textsc{Maximum Weight Vertex Cover} (MWVC), and \textsc{Maximum Weight Independent Set} (MWIS) problems with a focus on those using data reduction rules. For more details on data reduction techniques used on other problems, we refer the reader to the recent survey~\cite{Abu-Khzam2022}. The MWC, MWVC, and MWIS are complementary, meaning if we find an MWIS in a graph $G$, we simultaneously find an MWVC in $G$ and an MWC in $\overline{G}$. All of these problems have been extensively studied, and a large portfolio of exact algorithms and heuristics has been developed. Figure~\ref{fig:history} illustrates the rich history and how new solvers are continually compared across these problems. It also highlights the recent shift towards using data reductions for all three problems.

\subsection{Exact Methods}

Exact algorithms compute optimal solutions by systematically exploring the solution space. A frequently used paradigm in exact algorithms for combinatorial optimization problems is called \emph{branch-and-bound}~\cite{land1960automatic}. One of the earliest results using this technique for the problems we consider here was the MWC solver called \cliquer~\cite{ostergaard1999new}. In the following, we cover the exact solvers developed for the MVC, MWVC, and MWIS in that order.

Since \cliquer, several more branch-and-bound solvers for the MWC problem have been presented. These branch-and-bound solvers can broadly be placed in two categories. The first category uses \textsc{MaxSAT} reasoning to prune the search space and includes the two branch-and-bound algorithms called \mwclq{}~\cite{fang2016exact}, and \tsmmwc{}~\cite{jiang2018two}. The second category focuses on data reductions instead. It includes the \wlmc{}~\cite{jiang2017exact} and \mwcredu{}~\cite{erhardt2023improved} algorithms. The first algorithm \wlmc{} utilize a straightforward upper/lower bound reduction rule, where the heaviest known clique is used as a lower bound. Then, for any vertex $u$, an upper bound on the heaviest clique containing $u$ is $\textsc{UB}_0(u) = \w(N[u])$. If this upper bound is less than or equal to the lower bound, $u$ can be removed. In addition to the fast $\textsc{UB}_0$, they also consider a slightly more complicated upper bound that tries to exclude the heaviest neighbor. With the most recent algorithm, \mwcredu{}, Erhardt~\etal~introduced several new reduction rules that significantly improved the state-of-the-art exact solvers. These include reductions based on twins, domination, and simplicial vertices. In 2024, Rozman~\etal~\cite{rozman2024exact} introduced a branch-and-bound method called \maxcliqueweight{} for the MWC problem using simple bounds based on coloring. In their solver, the authors do not use data reduction rules, and in the experiments, \maxcliqueweight{} is compared only to the MWC solver \cliquer~\cite{ostergaard1999new}, which was presented in 2002.

For the MWVC and MWIS, only one recent exact solver called \sbms~\cite{xu2016new} did not utilize data reductions. Instead, \sbms{} uses a series of \textsc{SAT} formulations that each answered if there is an MWVC of a given size. Since \sbms{}, every exact algorithm presented for these two problems relies on reduction rules. The first in this sequence, \bmwvc~\cite{wang2019exact}, analyzed the effectiveness of the reductions and showed that reduction rules often reduce massive graphs \hbox{to tractable sizes.}

Such reduction rules have also been added to branch-and-bound methods yielding so-called \emph{branch-and-reduce} algorithms~\cite{akiba-tcs-2016}. These algorithms extend upon branch-and-bound by applying reduction rules to the current graph before each branching step. \kamis~\cite{lamm2019exactly} was the first branch-and-reduce solver introduced for these problems. It has since become a highly influential solver that introduced several new reduction rules. The authors first introduced two meta-reductions called neighborhood removal and neighborhood folding, from which they derived a new set of weighted reduction rules. On this foundation, a branch-and-reduce algorithm was developed using pruning with weighted clique covers similar to the approach by Warren and Hicks~\cite{warren2006combinatorial} for upper bounds and an adapted version of the ARW local search~\cite{andrade-2012} for lower bounds. The \kamis~algorithm was then extended to \struction{} by Gellner~\etal~\cite{gellner2021boosting} to utilize different struction based reduction rules that were originally introduced by Ebenegger~\etal~\cite{ebenegger1984pseudo} and later improved by Alexe~\etal~\cite{alexe2003struction}. In contrast to previous reduction rules, struction rules do not necessarily decrease the graph size but rather transform the graph, which can lead to further reduction. Two other exact solvers using the branch-and-reduce approach were also recently introduced, called \solve~\cite{xiao2021efficient} and \cbnr~\cite{liu2023application}. These solvers use more computationally expensive reduction \hbox{rules than \kamis.}

In a recent theoretical result, Xiao~\etal~\cite{xiao2024maximum} presented a branch-and-bound algorithm idea using reduction rules working especially well on sparse graphs. They perform a detailed analysis of the running time bound on special graphs in their theoretical work. With the measure-and-conquer technique, they show that the running time of their algorithm is~$\mathcal{O}^*(1.1443^{(0.624x-0.872)n})$ where $x$ is the average degree of the graph. This improves previous time bounds for this problem using polynomial space complexity for graphs of average degree \hbox{up to three.}
    
For the unweighted MVC problem, Figiel~\etal~\cite{DBLP:conf/esa/FigielFNN22}~introduced a new idea to the state-of-the-art way of applying reductions. They propose not only to perform reductions but also the possibility of undoing them during the reduction process. As they showed in their results, this can lead to new possibilities to apply further reductions and finally to \hbox{smaller reduced graphs.}

\subsection{Heuristic Methods}

A widely used heuristic approach is called \textit{local search}, which tries to improve any feasible solution by simple insertion, removal, or swap operations. Although, in theory, local search generally offers no guarantees for the quality of the solution, in practice, it routinely finds high-quality solutions significantly faster than exact procedures. Almost every heuristic for the MWC, MWVC, and MWIS problems is based on local search.

For unweighted graphs, the iterated local search (\textsc{ARW}) by Andrade~\etal~\cite{andrade-2012} was a very successful heuristic. It is based on so-called $(1,2)$-swaps, which remove one vertex from the solution and add two new vertices, thus improving the current solution by one. Their algorithm uses special data structures that find such a $(1,2)$-swap in $\mathcal{O}(m)$ time or prove that none exists.

Again, we cover the heuristics for these problems in the order of MVC, MWVC, and then MWIS. A central topic in local search for the MWC problem is how to escape from local optima. For the MWC, several techniques have been added to local search to address this, including tabu search used in \mnts~\cite{wu2012multi}, adaptive perturbation in \bls~\cite{benlic2013breakout}, configuration checking in \lsccbps~\cite{wang2016two}, smart restarts used in \rrwl~\cite{fan2017restart}, and walk perturbation in \textsc{SCCWalk} and \sccwalk~\cite{wang2020sccwalk}. The two solvers \textsc{ReTS1} and \rets~\cite{zhou2017push} also added a new \textit{push} operator that can simultaneously add and remove vertices from a solution, compared to the typical add and swap operators. As with exact methods, using data reductions in heuristics is also becoming more common. For the MWC problem, this was first introduced in the \fastwclq~\cite{cai2016fast} and later improved under the same name~\cite{cai2021semi}. We refer to the second version as \fastwclqn. These heuristics used the upper/lower bound reductions mentioned earlier. The most recent heuristic for the MWC problem, \mwcpeel~\cite{erhardt2023improved}, does not use local search but a technique called reduce-and-peel~\cite{chang2017} instead. This \textit{reduce-and-peel} is a greedy approach that uses exact reduction rules whenever possible. A heuristic tie-breaking mechanism is needed to ensure progress when exact reductions can no longer reduce the graph. The \mwcpeel{} was introduced alongside \mwcredu{} and used the same extensive \hbox{set of reductions.}

For the MWVC problem, the earliest heuristics used ant colony optimization. The first was called \aco~\cite{shyu2004ant}, which was later improved resulting in \acosee~\cite{jovanovic2011ant}. The next two heuristics used multi-start iterated tabu search~\cite{zhou2016multi} (\msits) and a population-based iterated greedy heuristic~\cite{bouamama2012population} (\pbig). Since then, a technique based on dynamic edge-weights has been widely adopted for the MWVC problem. The technique was first introduced in \dlswcc~\cite{li2016efficient} and has since been used by several heuristics. Subsequent iterations of this technique brought new improvements, starting with \fastwvc~\cite{cai2016fast} that added a construction procedure to generate a high-quality initial vertex cover. Then, \numwvc~\cite{li2020numwvc} added reduction rules as a preprocessing step to reduce the graph size. Two heuristics called \textsc{DynWVC} and \dynwvc~\cite{cai2018improving} introduced dynamic strategies for selecting which vertices to add or remove during the search. \maehts~\cite{wang2021fast} combined an evolutionary algorithm with reduction rules on top of the local search. The most recent heuristic to use this edge-weight technique is a hybrid method called \gnnvc~\cite{langedal2022efficient}. To construct the initial solution, \gnnvc{} combines data reductions and Graph Neural Networks in a reduce-and-peel approach. Two other recent heuristics deviate from this edge-weight technique. First, a population-based game-theoretic optimizer~\cite{qiu2022population} (\pgto), and second, an evolutionary algorithm based on the snowdrift game~\cite{li2023evolutionary} (\egmwvc). Neither of these last two heuristics \hbox{utilized data reductions.}

For the MWIS problem, a slightly different variation of local search has been frequently used, called iterated local search~\cite{lourencco2003iterated}. This metaheuristic makes random perturbations to the solution to escape local optima. Following the early results of \plswis~\cite{pullan2009optimisation}, the hybrid iterated local search heuristic \ilsvnd{} (often called \textsc{HILS}) by Nogueira~\etal~\cite{nogueira2018hybrid} adapted the \textsc{ARW} algorithm for weighted graphs. In addition to weighted $(1,2)$-swaps, it also uses $(\omega,1)$-swaps that add one vertex $v$ into the current solution and exclude its neighbors. Recently, the heuristic \metamis~\cite{dong2022metaheuristic} further improved on \ilsvnd{} by incorporating alternating augmenting-path moves.

The reduce-and-peel approach is also frequently used for the MWIS problem. Here, this method was first used in \dttwo~\cite{zheng2020efficient} and later improved resulting in \htwis~\cite{gu2021towards} and \hglv~\cite{tan2024efficient}. Another heuristic called \mmwis~\cite{grossmann2023finding} uses an elaborate version of reduce-and-peel, combining data reduction rules with an evolutionary approach. The authors perform heuristic reductions by utilizing information from the population that evolved during the evolution process. After performing this heuristic reduction, \mmwis{} return to exact reductions \hbox{as in reduce-and-peel.}

Another heuristic for the MWIS problem is called \bsa{} and is presented by Haller and Savchynskyy \cite{haller2024bregman}. This heuristic differed from the typical local search and reduce-and-peel heuristics presented earlier. Instead, Haller and Savenchynskyy introduced a Bregman-Sinkhorn Algorithm (\bsa) that addresses a family of clique cover LP relaxations. From the most recent heuristics, only \bsa{} and \metamis{} do not use reduction rules. These heuristics were evaluated on a newly published dataset of vehicle routing (VR) instances~\cite{dong2021new} that are exceptionally hard to reduce. These instances present a new challenge for practical data reductions.

In 2025, further work was done on the MWIS problem. First, the \dynls{} method was introduced by Zhu~\etal~\cite{zhu2025dynamic}. In their local search method, the authors use multiple strategies. Aside from a reduction preprocessing, they also have methods to accelerate convergence, escape local optima as well as a new variable neighborhood descent strategy combined with a reward mechanism to guide the search to high-quality solutions quickly.
Since more and more complex data reductions have been introduced and exhaustively applying these rules is computationally expensive, Gro{\ss}mann~\etal~\cite{grossmann2025accelerating} presented a graph neural network screening framework called \lnr{} that applies reduction rules only where they are predicted to succeed. With this approach, the authors show that it can be beneficial to also use these computationally expensive rules, even for heuristic methods, where fast running times are very important.
Another local-search-based method was introduced by Gro{\ss}mann~\etal~\cite{grossmann2025concurrent}. In this method, called \chils{}, multiple local search solutions are used to guide the search. Even though the method itself does not use reduction rules directly, the authors show in their experiments that the performance on most instances can be improved when combining \chils{} with \lnr{}. Only for the VR instances did the data \hbox{reductions not help.}

Recently, Borowitz~\etal~\cite{borowitz2026distributed} published distributed data reduction algorithms for the MWIS problem. Since the reductions introduced in their paper are based on the sequential reductions presented here, we will not specifically focus on distributed reductions \hbox{in this survey.}

\section{Data Reduction Rules}
\label{sec:reduction-rules}
This section documents the previously published data reduction rules for the MWIS problem. The reductions are grouped into different categories based on common properties. Each section starts with a brief introduction and an intuition for the presented rules. Note that the rules are not ordered by their complexity. The enumeration and labeling of the rules will not change, and new rules will be added at the end in the corresponding section. The reduction rules are presented using a standardized scheme shown in Reduction~\ref{red:example}.

\begin{reduction}[\lbrack Reduction Name\rbrack{} by \lbrack Authors\rbrack]\label{red:example}
    Description of the pattern that can be reduced.\\
    \reductiondetails{How to build the reduced graph $G'$}{Which weight can be added to the offset}{How to reconstruct the solution $\I$ for the original graph given the solution $\I'$ on the reduced graph $G'$}
\end{reduction}

First, we give the name of the reduction rule and cite the papers where the rule was first introduced. Then, we define the pattern that this rule can reduce. Finally, we give details on how to perform the actual reduction. This last information consists of three parts. First is information on constructing the reduced graph, called $G'$. Then, the \textit{offset} describes the difference between the weight of an MWIS on the reduced graph $\aw(G')$ and the weight of an MWIS on the original graph $\aw(G)$. Lastly, the information on how the solution on the reduced instance, called $\I'$, can be lifted to a solution on the original graph, called $\I$, \hbox{is provided.}

In addition to including or excluding vertices directly, some reduction rules combine multiple vertices into potentially new vertices. This combine procedure is called \textit{folding}. Including or excluding the folded vertices from the solution $\I$ only depends on whether the vertices they are folded into are included or excluded in the solution $\I'$ on the reduced instance. To be more precise, folding a set of vertices $X\subset V$ into a new vertex $v'$ generally results in a new graph $G'=G[V-X\cup \{v'\}]$, where the new vertex $v'$ is connected to all vertices in the neighborhood of $X$. If the set $X$ is folded into a vertex $v$ already existing in $G$, then the neighborhood is extended by the neighbors of $X$, \ie $N(v) = N(v)\cup N(X)$.

\subsection{Low Degree Reduction Rules}
\label{sec:deg-bounded-rules}
In this section we cover data reduction rules applicable to vertices with a specific degree. The presented rules fully cover all vertices of degree one and degree two. These are special cases of more powerful reductions presented \hbox{in later sections.}\newpage
\begin{reduction}[Degree One by Gu~\etal~\cite{gu2021towards} ]\phantomsection\label{red:deg1}
	Let $u,v\in V$ with $N(v) = \{u\}$.
	\begin{itemize}
		\item If $\w(v) \geq \w(u)$: include $v$. \\
        \reductiondetails{$G'=G-N[v]$ }{$\aw(G) = \aw(G') + \w(v)$}{$\I=\I'\cup\{v\}$}
	     \item If $\w(v) < \w(u)$: fold $u$ and $v$ into new vertex $v'$. \\
\reductiondetails{$G'=G[(V\cup\{v'\})\setminus\{u,v\}]$ with $N(v') = N(u)$ and $\w(v')=\w(u)-\w(v)$}{$\aw(G) = \aw(G') + \w(v)$}{If $v' \in \I'$, then $\I = \I' \setminus \{v'\}\cup \{u\}$, else $\I = \I' \cup \{v\}$}
        \end{itemize}
\end{reduction}

\begin{figreduction}[Triangle by Gu~\etal~\cite{gu2021towards}]{triangle}
	Let $v\in V$ be a degree-two vertex with two adjacent neighbors $x,y\in V$. Without loss of generality, assume ${\w(x) \leq \w(y)}$.
	\begin{itemize}
		\item If $\w(v) < \w(x)$: fold $v$ into $x$ and $y$.\\
        \reductiondetails{$G'=G-v$ and $\w(x) = \w(x) - \w(v),\ \w(y)  = \w(y) - \w(v)$}{$\aw(G) = \aw(G')+ \w(v)$}{If $x,\,y \notin \I'$, then $\I = \I' \cup \{v\}$, else $\I = \I'$}
        
		\item If $\w(x) \leq \w(v) < \w(y)$: exclude $x$ and fold $v$ into $y$.\\
        \reductiondetails{$G'=G-\{v,x\}$ and $\w(y)  = \w(y) - \w(v)$}{$\aw(G) = \aw(G')+ \w(v)$}{If $y \notin \I'$, then $\I = \I' \cup \{v\}$, else $\I = \I'$}
		\item If $\w(v) \geq \w(y)$:  include $v$.\\
        \reductiondetails{$G'=G-N[v]$}{$\aw(G) = \aw(G') + \w(v)$}{$\I = \I' \cup \{v\}$}
	\end{itemize}
\end{figreduction}
\begin{figure}
	\centering
	\tikzsetnextfilename{triangle.pdf}
	\begin{tikzpicture}
		\begin{scope}[xshift=4cm, yshift=4.5cm]
			\node[tnode=$v$] (v) at (5,4) {$v$};
			\node[] (vtmp) [below=0.5cm of v] {};
			\node[rnode=$y$] (y) [right=0.25cm of vtmp] {$y$};
			\node[lnode=$x$] (x) [left=0.25cm of vtmp] {$x$};
			\node[weight] (vweight) at ($(v) + (0,0.25)$) {$\w_v$};
			\node[weight] (wweight) at (x) {$\w_x$};
			\node[weight] (uweight) at (y) {$\w_y$};
			\remainingG{5,2}{$G-\{v,x,y\}$}
			\draw[edge] (y) -- (6,2);
			\draw[edge] (y) -- (5.5,2);
			\draw[edge] (y) -- (5,2);
			\draw[edge] (x) -- (5,2);
			\draw[edge] (x) -- (4.5,2);
			\draw[edge] (x) -- (4,2);
			\draw[edge] (y) -- (v);
			\draw[edge] (x) -- (v);
			\draw[edge] (y) -- (x);
			\draw[black,-{Stealth[scale=1.5]}] (3.5,1.5) -- (1.7,0);
			\draw[black,-{Stealth[scale=1.5]}] (6.5,1.5) -- (8.3,0);
			\draw[black,-{Stealth[scale=1.5]}] (5,1.5) -- (5,.2);
		\end{scope}
	    \node[weightText] (case0) at (5.5,5.5) {$\w_v < \w_x$};
		\node[nodeR] (v) at (5,4) {};
		\node[] (vtmp) [below=0.5cm of v] {};
		\node[rnode=$y$] (y) [right=0.25cm of vtmp] {};
		\node[lnode=$x$] (x) [left=0.25cm of vtmp] {};
		\remainingG{5,2}{$G-\{v,x,y\}$}
	    \node (offset) at (5,1) {$\aw(G) = \aw(G')+\w_v$};
		\draw[edge] (y) -- (6,2);
		\draw[edge] (y) -- (5.5,2);
		\draw[edge] (y) -- (5,2);
		\draw[edge] (x) -- (5,2);
		\draw[edge] (x) -- (4.5,2);
		\draw[edge] (x) -- (4,2);
		\draw[edgeR] (y) -- (v);
		\draw[edgeR] (x) -- (v);
		\draw[edge] (y) -- (x);
		\node[weight] (xweight) at ($(x) + (-.5,0)$) {$\w_x-\w_v$};
		\node[weight] (yweight) at ($(y) + (.5,0)$) {$\w_y-\w_v$};
%
	    \begin{scope}[xshift=4cm]
	    	\node[weightText] (casew) at (5,5.5) {$\w_x\leq\w_v<\w_y$};
	    	\node[nodeR] (v) at (5,4) {};
	    	\node[] (vtmp) [below=0.5cm of v] {};
	    	\node[rnode=$y$] (y) [right=0.25cm of vtmp] {};
	    	\node[nodeE] (x) [left=0.25cm of vtmp] {};
	    	\remainingG{5,2}{$G-\{u,x,y\}$}
	    	\node (offset) at (5,1) {$\aw(G) = \aw(G')+\w_v$};
	    	\draw[edge] (y) -- (6,2);
	    	\draw[edge] (y) -- (5.5,2);
	    	\draw[edge] (y) -- (5,2);
	    	\draw[edgeR] (x) -- (5,2);
	    	\draw[edgeR] (x) -- (4.5,2);
	    	\draw[edgeR] (x) -- (4,2);
	    	\draw[edgeR] (y) -- (v);
	    	\draw[edgeR] (x) -- (v);
	    	\draw[edgeR] (y) -- (x);
	    	\node[weight] (uweight) at (y) {$\w_y-\w_v$};
	    \end{scope}
	
	    \begin{scope}[xshift=8cm]
	    	\node[weightText] (case) at (4.5,5.5) {$\w_y\leq \w_v$};
	    	\node[nodeI] (v) at (5,4) {};
	    	\node[] (vtmp) [below=0.5cm of v] {};
	    	\node[nodeE] (x) [right=0.25cm of vtmp] {};
	    	\node[nodeE] (y) [left=0.25cm of vtmp] {};
	    	\remainingG{5,2}{$G-\{u,x,y\}$}
	    	\node (offset) at (5,1) {$\aw(G) = \aw(G')+\w_v$};
	    	\draw[edgeR] (x) -- (6,2);
	    	\draw[edgeR] (x) -- (5.5,2);
	    	\draw[edgeR] (x) -- (5,2);
	    	\draw[edgeR] (y) -- (5,2);
	    	\draw[edgeR] (y) -- (4.5,2);
	    	\draw[edgeR] (y) -- (4,2);
	    	\draw[edgeR] (x) -- (v);
	    	\draw[edgeR] (y) -- (v);
	    	\draw[edgeR] (x) -- (y);
	    \end{scope}
	\end{tikzpicture}
	\caption{Different cases of Reduction~\ref{red:triangle} with $\w_x\leq\w_y$. The status of a vertex after reducing is shown by its color, where green means included, red is excluded, and gray is folded.}\label{fig:triangle}
\end{figure}
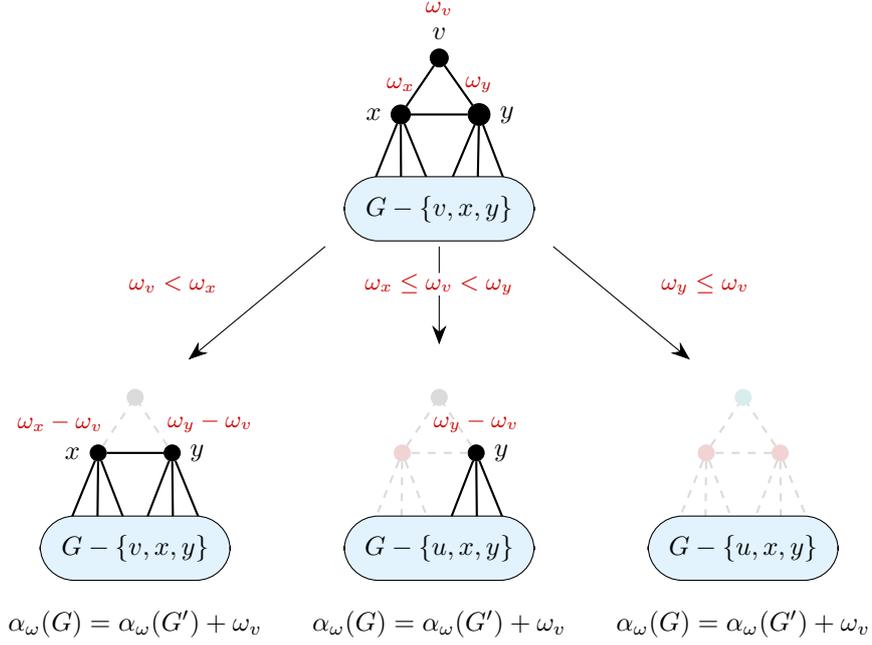 
\newpage	
\begin{figreduction} [V-Shape by Gu~\etal~\cite{gu2021towards} and Lamm~\etal~\cite{lamm2019exactly}]{vShape}
	Let $v\in V$ be a degree-two vertex with two non-adjacent neighbors $x,y \in V$. Without loss of generality, assume ${\w(x) \leq \w(y)}$.
	\begin{itemize}
		\item If $\w(v) < \w(x)$: fold $v$ into new vertex $v'$.\\
        \reductiondetails{$G'=G[(V\cup\{v'\})\setminus\{v\}]$ with ${N(v') = N(x) \cup N(y)}$ and set \hfill \break ${\w(x) = \w(x) - \w(v)}$, $\w(y) = \w(y) - \w(v)$}{$\aw(G) = \aw(G')+ \w(v)$}{If $x\in\I' or y\in \I'$, then $\I = \I' \setminus \{v\}$, else $\I = \I'$}
		\item If ${\w(x) \leq \w(v) < \w(y)}$: fold $v$ into $x$ and $y$.\\
        \reductiondetails{$G'=G-v$ with ${N(x) = N(x) \cup N(y)}$ and $\w(y) = \w(y) - \w(v)$}{$\aw(G) = \aw(G')+ \w(v)$}{If $x,y\notin \I'$, then $\I = \I' \cup \{v\}$, else $\I = \I'$}        
		\item If $\w(y) \leq \w(v)$ and $\w(x) + \w(y) \leq \w(v)$, include $v$.\\  
        \reductiondetails{$G'=G-N[v]$}{$\aw(G) = \aw(G') + \w(v)$}{$\I = \I' \cup \{v\}$}
		\item  If $\w(y) \leq \w(v)$ and $\w(x) + \w(y) > \w(v)$: fold $v$, $x$, $y$ into a new vertex $v'$.\\
        \reductiondetails{$G'=G[(V\cup\{v'\})\setminus\{v,\,x,\,y\}]$ with ${N(v') = N(x) \cup N(y)}$\\ & and $\w(v') = \w(x) + \w(y) - \w(v)$}{$\aw(G) = \aw(G')+ \w(v)$}{If $v'\in \I'$, then $\I = \I' \cup \{x,y\}\setminus \{v'\}$, else $\I = \I' \cup \{v\}\setminus \{v'\}$} 
	\end{itemize}
\end{figreduction}
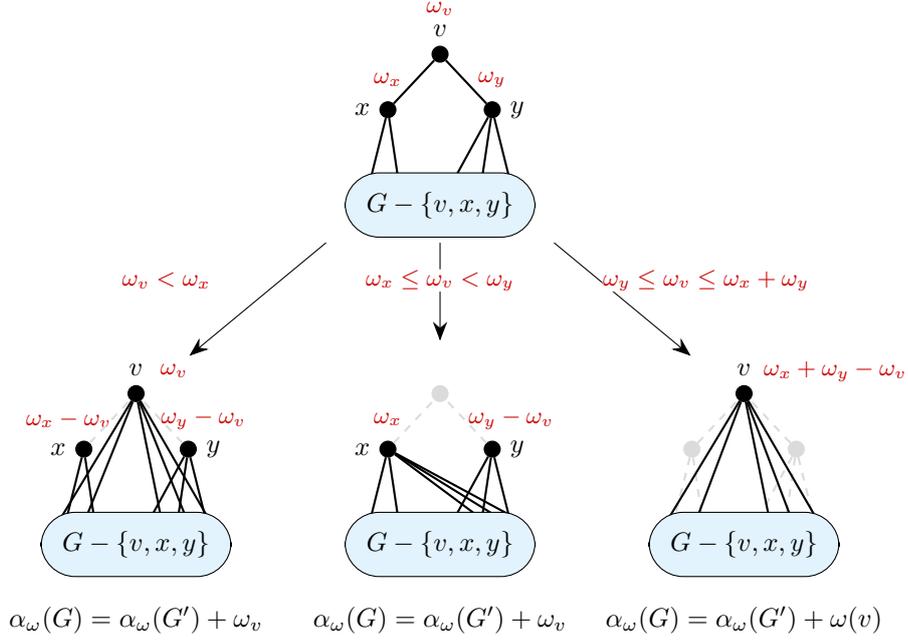
\begin{figure}
	\centering
	\tikzsetnextfilename{vShape.pdf}
	\begin{tikzpicture}
		\begin{scope}[xshift=4cm, yshift=4.5cm]
			\node[tnode=$v$] (v) at (5,4) {};
			\node[] (vtmp) [below=0.5cm of v] {};
			\node[rnode=$y$] (u) [right=0.45cm of vtmp]{};
			\node[lnode=$x$] (w) [left=0.45cm of vtmp]{};
			\node[weight] (vweight) at ($(v) + (0,0.2)$) {$\w_v$};
			\node[weight] (wweight) at (w) {$\w_x$};
			\node[weight] (uweight) at (u) {$\w_y$};
			\remainingG{5,2}{$G-\{v,x,y\}$}
			\draw[edge] (u) -- (6,2);
			\draw[edge] (u) -- (5.5,2);
			\draw[edge] (u) -- (5,2);
			\draw[edge] (w) -- (4.5,2);
			\draw[edge] (w) -- (4,2);
			\draw[edge] (u) -- (v);
			\draw[edge] (w) -- (v);
			\draw[black,-{Stealth[scale=1.5]}] (3.5,1.5) -- (1.7,0);
			\draw[black,-{Stealth[scale=1.5]}] (6.5,1.5) -- (8.3,0);
			\draw[black,-{Stealth[scale=1.5]}] (5,1.5) -- (5,.2);
		\end{scope}
	    \node[weightText] (case0) at (5.4,5.5) {$\w_v< \w_x$};
	    \node (offset) at (5,1) {$\aw(G) = \aw(G')+\w_v$};
	    \node[tnode=$v$] (v) at (5,4){};
		\node[] (vtmp) [below=0.5cm of v] {};
		\node[rnode=$y$] (u) [right=0.45cm of vtmp]{};
		\node[lnode=$x$] (w) [left=0.45cm of vtmp]{};
		\node[weight] (vweight) at ($(v) + (.5, -0.1)$) {$\w_v$};
		\node[weight] (wweight) at ($(w) + (-0.2, 0)$) {$\w_x - \w_v$};
		\node[weight] (uweight) at ($(u) + (0.2, 0)$) {$\w_y-\w_v$};
		\remainingG{5,2}{$G-\{v,x,y\}$}
		\draw[edge] (u) -- (6,2);
		\draw[edge] (u) -- (5.5,2);
		\draw[edge] (u) -- (5,2);
		\draw[edge] (w) -- (4.5,2);
		\draw[edge] (w) -- (4,2);
		\draw[edgeR] (u) -- (v);
		\draw[edgeR] (w) -- (v);
		\draw[edge] (v) -- (6.15,2);
		\draw[edge] (v) -- (5.8,2);
		\draw[edge] (v) -- (5.4,2);
		\draw[edge] (v) -- (4.2,2);
		\draw[edge] (v) -- (3.8,2);
%
	    \begin{scope}[xshift=4cm]
	    	\node[weightText] (casew) at (5,5.5) {$\w_x\leq\w_v<\w_y$};
	    	\node (offset) at (5,1) {$\aw(G) = \aw(G')+\w_v$};
	    	\node[nodeR] (v) at (5,4){};
	    	\node[] (vtmp) [below=0.5cm of v] {};
	    	\node[rnode=$y$] (u) [right=0.45cm of vtmp]{};
	    	\node[lnode=$x$] (w) [left=0.45cm of vtmp]{};
	    	\node[weight] (wweight) at (w) {$\w_x$};
	    	\node[weight] (uweight) at ($(u) + (0.25, 0)$) {$\w_y-\w_v$};
	    	\remainingG{5,2}{$G-\{v,x,y\}$}
	    	\draw[edge] (u) -- (6,2);
	    	\draw[edge] (u) -- (5.5,2);
	    	\draw[edge] (u) -- (5,2);
	    	\draw[edge] (w) -- (4.5,2);
	    	\draw[edge] (w) -- (4,2);
	    	\draw[edge] (w) -- (6,2);
	    	\draw[edge] (w) -- (5.7,2.4);
	    	\draw[edge] (w) -- (5.9,2.4);
	    	\draw[edgeR] (u) -- (v);
	    	\draw[edgeR] (w) -- (v);
	    \end{scope}
	
	    \begin{scope}[xshift=8cm]
	    	\node[weightText] (case) at (4.7,5.5) {$\w_y\leq \w_v \leq \w_x + \w_y$};
	    	\node[tnode=$v$] (v) at (5,4){};
	    	\node[] (vtmp) [below=0.5cm of v] {};
	    	\node[nodeR] (u) [right=0.45cm of vtmp]{};
	    	\node[nodeR] (w) [left=0.45cm of vtmp]{};
	    	\node[weight] (vweight) at ($(v) + (1.2, -0.1)$) {$\w_x + \w_y -\w_v$};
	    	\remainingG{5,2}{$G-\{v,x,y\}$}
	    	\node (offset) at (5,1) {$\aw(G) = \aw(G') +\w(v)$};
	    	\draw[edgeR] (u) -- (6,2);
	    	\draw[edgeR] (u) -- (5.5,2);
	    	\draw[edgeR] (u) -- (5,2);
	    	\draw[edgeR] (w) -- (4.5,2);
	    	\draw[edgeR] (w) -- (4,2);
	    	\draw[edgeR] (u) -- (v);
	    	\draw[edgeR] (w) -- (v);	    	
	    	\draw[edge] (v) -- (6.15,2);
	    	\draw[edge] (v) -- (5.75,2);
	    	\draw[edge] (v) -- (5.4,2);
	    	\draw[edge] (v) -- (4.25,2);
	    	\draw[edge] (v) -- (3.8,2);
	    \end{scope}
	\end{tikzpicture}
	\caption{Different folding cases of Reduction~\ref{red:vShape} with weights $\w_x\leq\w_y$. }\label{fig:vShape}
\end{figure} 

The following reductions deal with special patterns containing degree two vertices such as paths and cycles. 
\begin{reduction}[3-Path Reduction by Xiao~\etal~\cite{xiao2024maximum}]\phantomsection\label{red:path}
    Let $v_1 v_2 v_3 v_4$ be a 3-path such that $\deg(v_2) = \deg(v_3) = 2$ and $\w(v_1) \geq \w(v_2) \geq \w(v_3) \geq \w(v_4)$, then fold $v_2$ and $v_3$ into the path.\\
    \reductiondetails{$G' = G-\{v_2,\,v_3\}$, add the edge $\{v_1,\,v_4\}$ \\& and set $\w(v_1) = \w(v_1) + \w(v_3 ) - \w(v_2)$}{$\aw(G) = \aw(G') + \w(v_2)$}{If $v_1\in\I'$, then $\I=\I' \cup \{v_3\}$, else $\I = \I'\cup \{v_2\}$}
\end{reduction}
\begin{reduction}[4-Path Reduction by Xiao~\etal~\cite{xiao2024maximum}]\label{red:4path}
    Let $v_1 v_2 v_3 v_4 v_5$ be a 4-path such that $\deg(v_2) = \deg(v_3) = \deg(v_4) = 2$ and $\w(v_1) \geq \w(v_2) \geq \w(v_3) \leq \w(v_4) \leq \w(v_5)$, then fold $v_2$ and $v_4$ into the path. \\
    \reductiondetails{$G' = G-\{v_2,\,v_4\}$, add edges $\{v_1,\,v_3\}$ and $\{v_3,\,v_5\}$, and \\& set $\w(v_1) = \w(v_1) + \w(v_3 ) - \w(v_2)$ and $\w(v_5) = \w(v_5) + \w(v_3 ) - \w(v_4)$}{$\aw(G) = \aw(G') + \w(v_2) + \w(v_4) - \w(v_3)$}{If $v_3 \in \I'$, then $\I = \I' \setminus \{v_3\} \cup \{v_2,\, v_4\}$, \\ & else if $v_1 \in \I'$ and $v_5 \notin \I'$, then $\I = \I' \cup \{v_4\}$ \\ & else if $v_1 \notin \I'$ and $v_5 \in \I'$, then $\I = \I' \cup \{v_2\}$ \\ & else $\I = \I' \cup \{v_3\}$}
\end{reduction}
\begin{reduction}[4-Cycle Reduction by Xiao~\etal~\cite{xiao2024maximum}]\phantomsection\label{red:4cycle}
    Let $v_1 v_2 v_3 v_4$ be a 4-cycle such that $\deg(v_2) = \deg(v_3) = 2$ and
    $\w(v_1) \geq \w(v_2) \geq \w(v_3)$, then fold $v_2$ and $v_3$ into the cycle. \\
    \reductiondetails{$G' = G - \{v_2,\,v_3\}$ and $\w(v_1) = \w(v_1 ) + \w(v_3) - \w(v_2)$}{$\aw(G) = \aw(G') + \w(v_2)$}{If $v_1\in\I'$, then $\I=\I' \cup \{v_3\}$, else $\I = \I'\cup \{v_2\}$}
\end{reduction}
\begin{reduction}[5-Cycle Reduction by Xiao~\etal~\cite{xiao2024maximum}]\phantomsection\label{red:5cycle}
    Let $v_1 v_2 v_3 v_4 v_5$ be a 5-cycle such that $\deg(v_2) = \deg(v_3) = \deg(v_5)= 2$, $\min\{\deg(v_1), \deg(v_4)\}\geq 3$ and
    $\w(v_1) \geq \w(v_2) \geq \w(v_3) \leq \w(v_4)$. 
    \begin{itemize}
        \item If $\w(v_3) > \w(v_5)$, then fold $v_5$ into the cycle.\\
        \reductiondetails{$G'=G-v_5$ and for all $i \in \{1,2,3,4\}$, set $\w(v_i)= \w(v_i) - \w(v_5)$}{$\aw(G) = \aw(G')+2 \cdot \w(v_5)$}{If $v_1, v_4 \notin \I'$, then $\I = \I' \cup \{v_5\}$, else $\I = \I'$ \\[-.5em]}
        \item If $\w(v_3) \leq \w(v_5)$, then fold $v_2$ and $v_3$ into the cycle.\\        
        \reductiondetails{$G'=G-\{v_2,\,v_3\}$ and set $\w(v_1)= \w(v_1) - \w(v_2)$, $\w(v_4)= \w(v_4) - \w(v_3)$ and $\w(v_5)= \w(v_5) - \w(v_3)$}{$\aw(G) = \aw(G')+\w(v_2)+\w(v_3)$}{If $v_1,v_4\in\I'$, then $\I = \I'$, \\&
        else if $v_1\in \I'$ and $v_4\notin \I'$, then $\I = \I' \cup \{v_3\}$, \\&
        else if $v_1\notin \I'$ and $v_4\in \I'$, then $\I = \I' \cup \{v_2\}$, \\&
        else $\I = \I' \cup \{v_2\}$}
    \end{itemize}
\end{reduction}
\begin{reduction}[6-Cycle Reduction by Xiao~\etal~\cite{xiao2024maximum}]\phantomsection\label{red:6cycle}
    Let $v_1 v_2 v_3 v_4 v_5v_6$ be a 6-cycle such that $\deg(v_2) = \deg(v_3) = \deg(v_5)= \deg(v_6) = 2$, $\w(v_1) \geq \max\{\w(v_2), \w(v_6)\}$, $\w(v_4) \geq \max\{\w(v_3), \w(v_5)\}$ and  $\w(v_6) \geq \w(v_5)$. 
    \begin{itemize}
        \item If $\w(v_2) \geq \w(v_3)$, then fold $v_5$ and $v_6$ into the cycle.\\
        \reductiondetails{$G'=G-\{v_5,\, v_6\}$, set $\w(v_2)= \w(v_2) + \w(v_6)$ and  $\w(v_3)= \w(v_3) + \w(v_5)$}{$\aw(G) = \aw(G')$}{If $v_2\in \I'$, then $\I = \I' \cup \{v_6\}$, \\& else if $v_3\in \I'$, then $\I = \I'\cup \{v_5\}$, \\& else $\I=\I$. \\[-.5em]}
        \item Else, fold $v_6$ into the cycle.\\        
        \reductiondetails{$G'=G-v_6$, add the edge $\{v_1,v_5\}$ and set weights\\ &
        $\w(v_2)= \w(v_2) +\w(v_6)$,
        $\w(v_3)= \w(v_3) + \w(v_5)$ and \\& 
        $\w(v_5)= \w(v_6) + \w(v_3) - \max\{\w(v_2) + \w(v_6),\, \w(v_3) + \w(v_5)\}$}{$\aw(G) = \aw(G')$}{
        If $v_1,\,v_3\in \I'$, then $\I = \I' \cup \{v_5\}$, \\&
        else if $v_2,\,v_4 \in \I'$, then $\I = \I' \cup \{v_6\}$, \\&
        else if $v_1,\,v_4\in \I'$, then $\I=\I'$, \\&
        else $\I = (\I'\setminus \{v_2,\, v_5\}) \cup \{v_3,\,v_6\}$}
    \end{itemize}
\end{reduction}

The patterns reduced in reductions~\ref{red:path} to~\ref{red:6cycle}
can also be reduced by applying reductions~\ref{red:vShape} and~\ref{red:triangle} to the degree two vertices. Since reductions~\ref{red:path} to~\ref{red:4cycle} need to search for a more complicated patterns in the graph, it might be more interesting to only use the more general reductions dealing with degree two vertices in practical application. However, note that using reductions~\ref{red:path} to~\ref{red:4cycle} can result in different reduced instances.

\subsection{Neighborhood Rules}
\label{sec:neighborhood-rules}
This section presents reduction rules that are reducing a vertex $v$ based on estimating or solving the independent set weight in the neighborhood $N(v)$. These are special cases and extensions \hbox{derived from Reduction~\ref{red:neighborhood_meta}.}
\begin{reduction}[Heavy Vertex by Lamm~\etal~\cite{lamm2019exactly}]\phantomsection\label{red:neighborhood_meta}
	   Let $v\in V$ and $\w(v)\geq \aw(G[N(v)])$, then include $v$.\\
       \reductiondetails{$G'=G-N[v]$ }{$\aw(G) = \aw(G') + \w(v)$}{$\I=\I'\cup\{v\}$}
\end{reduction}
Reductions~\ref{red:neighborhood removal} and~\ref{red:clique_neighborhood} are using an estimate for $\aw(G[N(v)])$ to apply \hbox{the idea more efficiently.}
\begin{reduction}[Neighborhood Removal by Lamm~\etal~\cite{lamm2019exactly}]\phantomsection\label{red:neighborhood removal}
	   Let $v\in V$ and $\w(v)\geq \w(N(v))$, then include $v$.\\
       \reductiondetails{$G'=G-N[v]$ }{$\aw(G) = \aw(G') + \w(v)$}{$\I=\I'\cup\{v\}$}
\end{reduction}

While Reduction~\ref{red:neighborhood removal} uses a simple bound by summing the neighborhoods weight, this bound can be tightened by using a clique cover in the neighborhood $N(v)$ and summing over the maximum weight vertices per clique. This gives an upper bound to the optimal solution and results in Reduction~\ref{red:clique_neighborhood}.
\begin{reduction}[Clique Neighborhood Removal by Lamm~\etal~\cite{kamis,lamm2019exactly}]\label{red:clique_neighborhood}
Let $v\in V$ and $C$ be a set of disjoint cliques in $N(v)$. If $\w(v) \geq \sum_C \max \{ \w(x) \mid x \in C \}$ \hbox{include $v$}.\\
\reductiondetails{$G'=G-N[v]$ }{$\aw(G) = \aw(G') + \w(v)$}{$\I=\I'\cup\{v\}$}
\end{reduction}
In the following, reductions~\ref{red:neighborhood folding} and~\ref{red:generalized_fold} introduce further reduction possibilities for the \hbox{case of $\w(v)<\aw(G[N(v)])$.}
\begin{reduction}[Neighborhood Folding by Lamm~\etal~\cite{lamm2019exactly}]\phantomsection\label{red:neighborhood folding}
Let $v \in V$, and suppose that $N(v)$ is independent. If $\w(N(v)) > \w(v)$, but $\w(N(v)) - \min \{ \w(u) \mid u \in N(v) \} < \w(v)$, then fold $v$ and $N(v)$ into a new vertex $v'$.\\
\reductiondetails{$G'=G[(V\cup \{v'\})\setminus N[v]]$ with $N(v')=N(N(v))$ and $\w(v') = \w(N(v)) - \w(v)$}{$\aw(G) = \aw(G') + \w(v)$}{If $v'\in \I'$, then $\I = (\I'\setminus\{v'\}) \cup N(v)$, else $\I = \I' \cup \{v\}$}    
\end{reduction}

The more general form of reducing a vertex $v$ and its neighborhood, is described in Reduction~\ref{red:generalized_fold}. For this reduction rule potentially multiple independent set problems have to be solved in the neighborhood, making the \hbox{rule very expensive.}
\begin{reduction}[Generalized Neighborhood Folding by Lamm~\etal~\cite{kamis, lamm2019exactly}] \label{red:generalized_fold}
	Let $v \in V$, then
    \begin{itemize}
    \item if $G[N(v)]$ contains only one independent set $\tilde{\I}$ with $\w(\tilde{\I}) > \w(v)$, fold $v$ and $N(v)$ into a new vertex $v'$.\\
    \reductiondetails{$G'=G[(V\cup \{v'\})\setminus N[v]]$ with $N(v')=N(N(v))$ and $\w(v') = \w(\tilde{\I}) - \w(v)$}{$\aw(G) = \aw(G') + \w(v)$}{If $v'\in \I'$, then $\I = (\I'\setminus\{v'\}) \cup \tilde{\I}$, else $\I = \I' \cup \{v\}$ \\[-.5em]} 
    \item if for $u\in N(v)$ all independent sets in $G[N(v)]$ including $u$ have less weight than $\w(v)$, exclude $u$.\\
    \reductiondetails{$G'=G-u$}{$\aw(G) = \aw(G')$}{$\I = \I'$ \\[-.5em]}
    \end{itemize}
\end{reduction}
Zheng~\etal~\cite{zheng2020efficient} introduced a 2-Vertex Neighbor Removal, an extension of Reduction~\ref{red:neighborhood removal} to two \hbox{non-adjacent vertices.}

\begin{reduction}[Two Vertex Neighborhood Removal by Zheng~\etal~\cite{zheng2020efficient}]\phantomsection\label{red:2vertex neighborhood removal}
	   Let $u,v\in V$ be non-adjacent and $\w(u)+\w(v)\geq \w(N(u)\cup N(v))$. Further assume for all vertices $x\in V$ $\w(x)< \w(N(x)$ (\ie Reduction~\ref{red:neighborhood removal} was applied), then include $u$ and $v$.\\
       \reductiondetails{$G'=G-N[\{u,\,v\}]$ }{$\aw(G) = \aw(G') +\w(u) + \w(v)$}{$\I=\I'\cup\{u,\,v\}$}
\end{reduction}
Xiao~\etal~\cite{xiao2021efficient} have further generalized the idea of Reduction~\ref{red:2vertex neighborhood removal} in Reduction~\ref{red:heavy_set} where they tighten the bound for these vertices further.
\begin{reduction}[Heavy Set by Xiao~\etal~\cite{xiao2021efficient}]\label{red:heavy_set}
	Let $u$ and $v$ be non-adjacent vertices having at least one
	common neighbor $x$. If for every independent set $\tI$ in the induced subgraph $G[N(\{u,\, v\})]$, $\w(N(\tI) \cap \{u,v\})\geq \w(\tI)$, then include $u$ and $v$.\\
    \reductiondetails{$G'=G-N[\{u, v\}]$}{$\aw(G) = \aw(G')+ \w(u) + \w(v)$}{$\I = \I' \cup \{u,v\}$}
\end{reduction}
\begin{remark}
    In \cite{xiao2021efficient} Reduction~\ref{red:heavy_set} for a heavy sets $\{u,\,v\}$ is only used if the neighborhood is small, \hbox{\ie $|N(\{u,v\})| \leq 8$.}
\end{remark}


\begin{reduction}[Heavy Set 3 by Gro{\ss}mann~\etal~\cite{grossmann2025accelerating}]\phantomsection\label{red:heavy_set3}
    Let $u$, $v$ and $w$ be vertices forming a heavy set, then include $u$, $v$ and $w$.\\
    \reductiondetails{$G' = G-N[\{u, v, w\}]$}{$\aw(G) = \aw(G') + \w(u) + \w(v)+ \w(w)$}{$\I = \I' \cup \{u,v,w\}$}
\end{reduction}

\subsection{Clique Based Rules}
\label{sec:clique-rules}
The following reductions are based on the observation that in a clique at most one vertex can be part of a maximum weight independent set.
A vertex $v$ where the neighborhood $N(v)$ forms a clique is called \emph{simplicial}. The first rule in this section works with these vertices.
\begin{reduction}[Simplicial Vertex by Lamm~\etal~\cite{lamm2019exactly}]\phantomsection\label{red:clique}
    Let $v\in V$ be simplicial with maximum weight in its neighborhood, \ie $\w(v)\geq \max \{\w(u) \mid u \in N(v)\}$, then include $v$.\\
    \reductiondetails{$G'=G-N[v]$}{$\aw(G) = \aw(G')+ \w(v)$}{$\I = \I' \cup \{v\}$}
\end{reduction}
\begin{reduction}[Simplicial Weight Transfer by Lamm~\etal~\cite{lamm2019exactly}] \label{red:clique_weight_transfer}
    Let $v\in V$ be simplicial, let $S(v)\subseteq N(v)$ be the set of all simplicial vertices. Further, let $\w(v) \geq \w(u)$ for all $u\in S(v)$. 
    \begin{itemize}
    \item If $\w(v)\geq \max \{\w(u) \mid u \in N(v)\}$, then use Reduction~\ref{red:clique}. 
    \item Else, fold $v$ into $N(v)$. \\
    \reductiondetails{Construct $G'$ by removing $v$ and all neighbors $u\in N(v)$ with $\w(u)\leq \w(v)$. Additionally,
	set the weight of all remaining neighbors ${x\in N(v)}$ to ${\w(x) = \w(x) - \w(v)}$}{$\aw(G) = \aw(G')+ \w(v)$}{If $\I' \cap N(v) = \emptyset$, then $\I = \I'\cup\{v\}$, else $\I = \I'$}	
    \end{itemize}
\end{reduction}

In the next reduction, almost simplicial vertices are considered. We call the pattern of the two vertices $u,v\in V$ with $u\in N(v)$ and $N(v)\setminus \{u\}$ forming a clique a $u$-$v$-\textit{funnel}.

\begin{figreduction}[Weighted Funnel by Gro{\ss}mann~\etal~\cite{grossmann2025accelerating}]{wFunnel}
    Assume $u,v \in V$ forms a $u$-$v$-funnel and that $\w(v) \geq \max\{\w(x) \mid x \in N(v)\setminus \{u\}\}$. Furthermore, let $N'(v) = \{x \in N(v) \setminus N[u] \mid \w(x) + \w(u) > \w(v)\}$.
    \begin{itemize}
        \item If $\w(v) \geq \w(u)$, fold $v$ and $u$ into its neighborhood.\\
              \reductiondetails{$G' = G - (N[v] \setminus N'(v))$ and for all $x \in N'(v)$, let $N(x) = N(x) \cup N(u)$ and \hbox{$\w(x) = \w(x) + \w(u) - \w(v)$}}{$\aw(G) = \aw(G') + \w(v)$}{If $\I' \cap N(v) = \emptyset$, then $\I = \I' \cup \{v\}$, else $\I = \I' \cup \{u\}$\\[-.5em]}
        \item If $\w(v) < \w(u)$, fold $v$ into its neighborhood.\\
              \reductiondetails{$G' = G - (N[v] \setminus (N'(v)\cup \{u\}))$ with $\w(u) = \w(u) - \w(v)$ and for all $x \in N'(v)$, let $N(x) = N(x) \cup N(u)$}{$\aw(G) = \aw(G') + \w(v)$}{If $\I' \cap N(v) = \emptyset$, then $\I = \I' \cup \{v\}$, else $\I = \I'$}
    \end{itemize}
\end{figreduction}

\begin{figure*}[t]
    \centering
    	\begin{tikzpicture}[scale=0.85]

		\node[rnode=$v$] (v) at (5.5,4.25) {};
		\node[below=0.5cm of v] (vtmp) {};
		\node[rnode=$x$] (x) [right=0.25cm of vtmp] {};
		\node[lnode=$w$] (w) [left=0.25cm of vtmp] {};
		\node[lnode=$u$] (u) [left=1cm of v] {};
		\remainingG{5,2}{$\quad G-N[v]\quad$}
		\draw[edge] (x) -- (6.15,2);
		\draw[edge] (x) -- (5.825,2);
		\draw[edge] (x) -- (5.5,2);
		\draw[edge] (w) -- (5.5,2);
		\draw[edge] (w) -- (5.175,2);
		\draw[edge] (w) -- (4.85,2);
		\draw[edge] (x) -- (v);
		\draw[edge] (w) -- (v);
		\draw[edge] (x) -- (w);
		\draw[edge] (u) -- (v);
		\draw[edge] (u) -- (w);
		\draw[edge] (u) -- (4.5,2);
		\draw[edge] (u) -- (4.25,2);
		\draw[edge] (u) -- (4,2);
		\node[weight] (vweight) at (v) {$21$};
		\node[weight] (uweight) at (u) {$20$};
		\node[weight] (wweight) at (w) {$2$};
		\node[weight] (xweight) at (x) {$4$};
		\draw[black,-{Stealth[scale=1.5]}] (6.5,3) -- (7.5,3) ;	
	    \begin{scope}[xshift=4cm]
			\node[nodeR] (v) at (5.5,4.25) {};
			\node[below=0.5cm of v] (vtmp) {};
			\node[node=$x$] (x) [right=0.25cm of vtmp] {};
			\node[nodeE] (w) [left=0.25cm of vtmp] {};
			\node[nodeR=$u$] (u) [left=1cm of v] {};
    		\remainingG{5,2}{$\quad G-N[v]\quad$}
			\draw[edge] (x) -- (6.15,2);
			\draw[edge] (x) -- (5.825,2);
			\draw[edge] (x) -- (5.5,2);
			\draw[edgeR] (w) -- (5.5,2);
			\draw[edgeR] (w) -- (5.175,2);
			\draw[edgeR] (w) -- (4.85,2);
			\draw[edgeR] (x) -- (v);
			\draw[edgeR] (w) -- (v);
			\draw[edgeR] (x) -- (w);
			\draw[edgeR] (u) -- (v);
			\draw[edgeR] (u) -- (w);
			\draw[edgeR] (u) -- (4.5,2);
			\draw[edgeR] (u) -- (4.25,2);
			\draw[edgeR] (u) -- (4,2);
		      \draw[edge] (x) -- (3.65,2.25);
		      \draw[edge] (x) -- (3.6,2.1);
		      \draw[edge] (x) -- (3.7,2);
			\node[weight] (xweight) at (x) {$3$};
	    \end{scope}
     \begin{scope}[xshift=8.5cm]
         
		\node[rnode=$v$] (v) at (5.5,4.25) {};
		\node[below=0.5cm of v] (vtmp)  {};
		\node[rnode=$x$] (x) [right=0.25cm of vtmp] {};
		\node[lnode=$w$] (w) [left=0.25cm of vtmp] {};
		\node[lnode=$u$] (u) [left=1cm of v] {};
		\remainingG{5,2}{$\quad G-N[v]\quad$}
		\draw[edge] (x) -- (6.15,2);
		\draw[edge] (x) -- (5.825,2);
		\draw[edge] (x) -- (5.5,2);
		\draw[edge] (w) -- (5.5,2);
		\draw[edge] (w) -- (5.175,2);
		\draw[edge] (w) -- (4.85,2);
		\draw[edge] (x) -- (v);
		\draw[edge] (w) -- (v);
		\draw[edge] (x) -- (w);
		\draw[edge] (u) -- (v);
		\draw[edge] (u) -- (w);
		\draw[edge] (u) -- (4.5,2);
		\draw[edge] (u) -- (4.25,2);
		\draw[edge] (u) -- (4,2);
		\node[weight] (vweight) at (v) {$21$};
		\node[weight] (uweight) at (u) {$22$};
		\node[weight] (wweight) at (w) {$2$};
		\node[weight] (xweight) at (x) {$4$};
		\draw[black,-{Stealth[scale=1.5]}] (6.5,3) -- (7.5,3);

	    \begin{scope}[xshift=4cm]
			\node[nodeR] (v) at (5.5,4.25) {};
			\node[below=0.5cm of v] (vtmp)  {};
			\node[rnode=$x$] (x) [right=0.25cm of vtmp] {};
			\node[nodeE] (w) [left=0.25cm of vtmp] {};
			\node[lnode=$u$] (u) [left=1cm of v] {};
    		\remainingG{5,2}{$\quad G-N[v]\quad$}
			\draw[edge] (x) -- (6.15,2);
			\draw[edge] (x) -- (5.825,2);
			\draw[edge] (x) -- (5.5,2);
			\draw[edgeR] (w) -- (5.5,2);
			\draw[edgeR] (w) -- (5.175,2);
			\draw[edgeR] (w) -- (4.85,2);
			\draw[edgeR] (x) -- (v);
			\draw[edgeR] (w) -- (v);
			\draw[edgeR] (x) -- (w);
			\draw[edgeR] (u) -- (v);
			\draw[edgeR] (u) -- (w);
			\draw[edge] (u) -- (4.5,2);
			\draw[edge] (u) -- (4.25,2);
			\draw[edge] (u) -- (4,2);
			\draw[edge] (x) -- (3.65,2.25);
			\draw[edge] (x) -- (3.6,2.1);
			\draw[edge] (x) -- (3.7,2);
			\node[weight] (xweight) at (x) {$4$};
			\node[weight] (uweight) at (u) {$1$};
	    \end{scope}
     \end{scope}
	\end{tikzpicture}
    \caption{Illustration from~\cite{grossmann2025accelerating} showing the Weighted Funnel (Reduction~\ref{red:wFunnel}), with $\w(v)\geq \w(u)$ left and $\w(v)<\w(u)$ right.
    }\label{fig:wFunnel}
\end{figure*}

\begin{remark}
    In the first case of Reduction~\ref{red:wFunnel}, where $\omega(v) \geq \omega(u)$, vertices $x\in N'(v)$ can be directly excluded, if $\omega(v)\geq \omega(u) + \omega(x)$.
\end{remark}

\subsection{Domination Based Rules}
\label{sec:dom-rules}

The following rules always compare the relation between two adjacent vertices and their neighborhood. In Reduction~\ref{red:dom} and~\ref{red:bse}, a vertex $v$ can be removed since it can always be replaced with its neighbor $u$ in an MWIS.
\begin{reduction}[Domination by Lamm~\etal~\cite{lamm2019exactly}]\phantomsection\label{red:dom}
	Let $u,\,v\in V$ be adjacent vertices such that $N[u]\subseteq N[v]$. If $\w(v)\leq \w(u)$, then exclude $v$.\\
    \reductiondetails{$G'=G-v$}{$\aw(G) = \aw(G')$}{$\I = \I'$} 
\end{reduction}
\newpage
\begin{figreduction}[Basic Single-Edge by Gu~\etal~\cite{gu2021towards}]{basicSE}\phantomsection\label{red:bse}
	Let $u,\,v\in V$ be adjacent vertices with $\w(N(u) \setminus N(v)) \leq\w(u)$, then exclude $v$.\\
    \reductiondetails{$G'=G-v$}{$\aw(G) = \aw(G')$}{$\I = \I'$}
\end{figreduction}
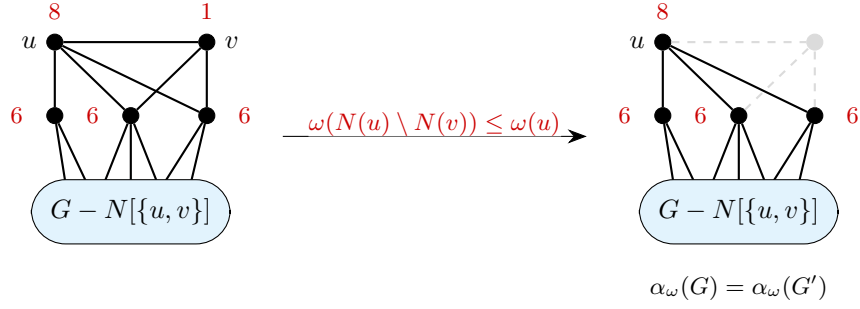
\begin{figure}
	\centering
	\tikzsetnextfilename{basicSE.pdf}
	\begin{tikzpicture}
		\node[rnode=$v$] (v) at (6,4.25) {};
		\node[lnode=$u$] (u) at ([shift={(-2,0)}]v) {};
		\node[node] (v3) [below=.75cm of v] {};
		\node[node] (v1) [below=.75cm of u] {};
		\node[node] (v2) at ($(v1)!0.5!(v3)$) {};
		\remainingG{5,2}{$G-N[\{u,v\}]$}
		\draw[edge] (u) -- (v);
		\draw[edge] (v3) -- (v);
		\draw[edge] (v2) -- (v);
		\draw[edge] (u) -- (v3);
		\draw[edge] (u) -- (v1);
		\draw[edge] (u) -- (v2);
		\draw[edge] (4.2,2) -- (v1);
		\draw[edge] (4.6,2) -- (v1);
		\draw[edge] (4.5,2) -- (v2);
		\draw[edge] (5,2) -- (v2);
		\draw[edge] (5.5,2) -- (v2);
		\draw[edge] (5.2,2) -- (v3);
		\draw[edge] (5.7,2) -- (v3);
		\node[weight] (vweight) at (v) {$1$};
		\node[weight] (uweight) at (u) {$8$};
		\node[lweight] (v1weight) at (v1) {$6$};
		\node[lweight] (v2weight) at (v2) {$6$};
		\node[rweight] (v3weight) at (v3) {$6$};
		\draw[black,-{Stealth[scale=1.5]}] (7,3) -- (11,3) node[weightColor, midway, above] {$\w(N(u)\setminus N(v)) \leq \w(u)$};
	    \begin{scope}[xshift=8cm]
	    	\node[nodeR] (v) at (6,4.25) {};
	    	\node[lnode=$u$] (u) at ([shift={(-2,0)}]v) {};
	    	\node[node] (v3) [below=.75cm of v] {};
	    	\node[node] (v1) [below=.75cm of u] {};
	    	\node[node] (v2) at ($(v1)!0.5!(v3)$) {};
	    	\remainingG{5,2}{$G-N[\{u,v\}]$}
	    	\node (offset) at (5,1) {\small$\aw(G) = \aw(G')$};
	    	\draw[edgeR] (u) -- (v);
	    	\draw[edgeR] (v3) -- (v);
	    	\draw[edgeR] (v2) -- (v);
	    	\draw[edge] (u) -- (v3);
	    	\draw[edge] (u) -- (v1);
	    	\draw[edge] (u) -- (v2);
	    	\draw[edge] (4.2,2) -- (v1);
	    	\draw[edge] (4.6,2) -- (v1);
	    	\draw[edge] (4.5,2) -- (v2);
	    	\draw[edge] (5,2) -- (v2);
	    	\draw[edge] (5.5,2) -- (v2);
	    	\draw[edge] (5.2,2) -- (v3);
	    	\draw[edge] (5.7,2) -- (v3);
	    	\node[weight] (uweight) at (u)  {\small$8$};
	    	\node[lweight] (v1weight) at (v1)  {\small$6$};
	    	\node[lweight] (v2weight) at (v2)  {\small$6$};
	    	\node[rweight] (v3weight) at (v3)  {\small$6$};
	    \end{scope}
	\end{tikzpicture}
	\caption{Illustration for Basic Single Edge; see Reduction~\ref{red:basicSE}.}\label{fig:basicSE}
\end{figure}

In contrast to the previous reductions in this section, Reduction~\ref{red:ese} covers the case where one of the two vertices $u$ or $v$ have to be in the solution.
\begin{figreduction}[Extended Single-Edge by Gu~\etal~\cite{gu2021towards}]{extendedSE}\phantomsection\label{red:ese}
	Let $u,\,v\in V$ be adjacent vertices with $\w(v) \geq \w(N(v)\setminus \{u\})$, then exclude $N(u) \cap N(v)$.\\
    \reductiondetails{$G'=G-(N(u) \cap N(v))$}{$\aw(G) = \aw(G')$}{$\I = \I'$}
\end{figreduction}

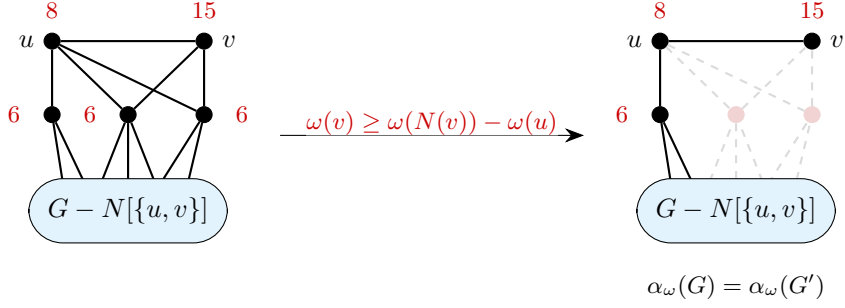
\begin{figure}[t]
	\centering
	\tikzsetnextfilename{extendedSE.pdf}
	\begin{tikzpicture}

	\node[rnode=$v$] (v) at (6,4.25) {};
	\node[lnode=$u$] (u) at ([shift={(-2,0)}]v) {};
	\node[node] (v3) [below=.75cm of v] {};
	\node[node] (v1) [below=.75cm of u] {};
	\node[node] (v2) at ($(v1)!0.5!(v3)$) {};
		\remainingG{5,2}{$G-N[\{u,v\}]$}
		\draw[edge] (u) -- (v);
		\draw[edge] (v3) -- (v);
		\draw[edge] (v2) -- (v);
		\draw[edge] (u) -- (v3);
		\draw[edge] (u) -- (v1);
		\draw[edge] (u) -- (v2);
		\draw[edge] (4.2,2) -- (v1);
		\draw[edge] (4.6,2) -- (v1);
		\draw[edge] (4.5,2) -- (v2);
		\draw[edge] (5,2) -- (v2);
		\draw[edge] (5.5,2) -- (v2);
		\draw[edge] (5.2,2) -- (v3);
		\draw[edge] (5.7,2) -- (v3);
		\node[weight] (vweight) at (v) {\small$15$};
		\node[weight] (uweight) at (u) {\small$8$};
		\node[lweight] (v1weight) at (v1) {\small$6$};
		\node[lweight] (v2weight) at (v2) {\small$6$};
		\node[rweight] (v3weight) at (v3) {\small$6$};
\draw[black,-{Stealth[scale=1.5]}] (7,3) -- (11,3) node[weightColor,midway, above] {\small$\w(N(v)\setminus\{u\})\leq \w(v)$};
\begin{scope}[xshift=8cm]
			\node[rnode=$v$] (v) at (6,4.25) {};
			\node[lnode=$u$] (u) at ([shift={(-2,0)}]v) {};
			\node[nodeE] (v3) [below=.75cm of v] {};
			\node[node] (v1) [below=.75cm of u] {};
			\node[nodeE] (v2) at ($(v1)!0.5!(v3)$) {};
	    	\remainingG{5,2}{$G-N[\{u,v\}]$}
	    	\node (offset) at (5,1) {\small$\aw(G) = \aw(G')$};
	    	\draw[edge] (u) -- (v);
	    	\draw[edgeR] (v3) -- (v);
	    	\draw[edgeR] (v2) -- (v);
	    	\draw[edgeR] (u) -- (v3);
	    	\draw[edge] (u) -- (v1);
	    	\draw[edgeR] (u) -- (v2);
	    	\draw[edge] (4.2,2) -- (v1);
	    	\draw[edge] (4.6,2) -- (v1);
	    	\draw[edgeR] (4.5,2) -- (v2);
	    	\draw[edgeR] (5,2) -- (v2);
	    	\draw[edgeR] (5.5,2) -- (v2);
	    	\draw[edgeR] (5.2,2) -- (v3);
	    	\draw[edgeR] (5.7,2) -- (v3);
	    	\node[weight] (vweight) at (v) {\small$15$};
	    	\node[weight] (uweight) at (u) {\small$8$};
	    	\node[lweight] (v1weight) at (v1) {\small$6$};
	    \end{scope}
	\end{tikzpicture}
	\caption{Illustration for Extended Single Edge; see Reduction~\ref{red:extendedSE}.}\label{fig:extendedSE}
\end{figure} 

As in Reduction~\ref{red:dom}, the following two rules are applied to two vertices $u$ and $v$ where $N(u) \subseteq N(v)$. Both of these rules are complementary: the first describes a pattern in which an edge can be removed, while the other reverses this reduction and shows a pattern in which an edge can be added. Depending on the method, both of these rules \hbox{can be useful.}

\begin{reduction}[Extended Domination by Gro{\ss}mann~\etal~\cite{grossmann2025accelerating}]\phantomsection\label{red:e_dom}
    Let $u, v\in V$ be adjacent vertices such that $N[u] \subseteq N[v]$. If $\w(v) > \w(u)$, then remove the edge between $u$ and $v$.\\
    \reductiondetails{$G'=(V, E \setminus \{u, v\})$ and \vfill \break $\w(v) = \w(v) - \w(u)$}{$\aw(G) = \aw(G')$}{If $v \in \I'$, then $\I = \I'\setminus \{u\}$, else $\I = \I'$}
\end{reduction}

\begin{reduction}[Extended Domination Reversed by Gro{\ss}mann~\etal~\cite{grossmann2025accelerating}]\phantomsection\label{red:e_dom_rev}
    Let $u, v\in V$ be non-adjacent vertices such that $N(u) \subseteq N(v)$. If $\w(u) + \w(v) < \w(N(v))$, then we can add an edge between $u$ and $v$.\\
    \reductiondetails{$G'=(V, E \cup \{u, v\})$ and \vfill \break $\w(v) = \w(v) + \w(u)$}{$\aw(G) = \aw(G')$}{If $v \in \I'$, then $\I = \I' \cup \{u\}$, else $\I = \I'$}
\end{reduction}

\subsection{Struction Based Rules}
\label{sec:struction-based-rules}
The weighted stability number data reduction rule, see Reduction~\ref{red:struction_original}, called struction was first introduced by Ebenegger~\etal~\cite{ebenegger1984pseudo} for the unweighted problem. All of these struction variants reduce the stability number $\aw(G)$ by the weight of the center vertex $v$ which the rule is applied to. Note that these rules can increase the graph size. We refer to this process as \emph{transforming}.
An important concept used in those reduction rules is \textit{layering}. For a given set $M$ of vertices $v_{x,y}$ with two indices $x\in X$ and $y\in Y$ a \textit{layer} $L_k= \{v_{x,y}\in M \mid x = k \}$ contains all elements with the first index equal to $k$. Note that in Reduction~\ref{red:struction_extended_reduced} the sets $X$ and $Y$ can contain vertices or vertex sets.  
\begin{reduction}[Original Weighted  Struction by Gellner~\etal~\cite{gellner2021boosting}]\phantomsection\label{red:struction_original}
    Let $v\in V$ such that $\w(v) = \min\{\w(u) \mid u \in N[v]\}$, then transform $v$. \\ 
    \reductiondetails{Construct the graph $G'$ as follows
    \begin{itemize}
        \item remove $v$ and set $\w(u) = \w(u) - \w(v)$ for each $u \in N(v)$
        \item for all $x,y\in N(v)$, if ${\{x,\, y\} \notin E}$ and ${x < y}$, then add a vertex $v_{x,y}$ with $N(v_{x,\,y})=N(\{x,\,y\})\setminus\{v\}$ and $\w(v_{x,y}) = \w(v)$
        \item for each $q\in N(v)$ and all $v_{q,x},v_{q,y}\in L_q$, if $\{x,\,y\}\in E$, then add the edge $\{v_{q,x}, \, v_{q,y}\}$                 
        \item for all $q,r\in N(v)$ with $q\neq r$ and all $v_{q,x}\in L_q$ and $v_{r,y}\in L_r$, add the edge $\{v_{q,x}, \, v_{r,y}\}$
    \end{itemize}}{$\aw(G) = \aw(G') + \w(v)$}{
    If $\I'\cap N(v) = \emptyset$ then $\I=\I' \cup \{v\}$, else $\I = \I' \cap V$} 
\end{reduction}

Gellner~\etal~\cite{gellner2021boosting} created further reductions based on Reduction~\ref{red:struction_original}. The first reduction rule they propose is a modification of Reduction~\ref{red:struction_original} such that the number of vertices in the solution $\I'$ is the same as in the original graph. This is achieved by assigning different weights and inserting additional edges, resulting in Reduction~\ref{red:struction_modified}.
\newpage
\begin{reduction}[Modified Weighted Struction by Gellner~\etal~\cite{gellner2021boosting}]\phantomsection\label{red:struction_modified}
Let $v\in V$ such that $\w(v) = \min\{\w(u) \mid u \in N[v]\}$, then transform $v$. \\ 
\reductiondetails{Construct the graph $G'$ as follows
    \begin{itemize}
        \item remove $v$ and set $\w(u) = \w(u) - \w(v)$ for each $u \in N(v)$
        \item for all $x,y\in N(v)$, if ${\{x,\, y\} \notin E}$ and ${x < y}$, then add a vertex $v_{x,y}$ with $N(v_{x,\,y})=N(\{x,\,y\})\setminus\{v\}$ and $\w(v_{x,y}) = \w(y)$
        \item for each $q\in N(v)$ and all $v_{q,x},v_{q,y}\in L_q$, if $\{x,\,y\}\in E$, then add the edge $\{v_{q,x}, \, v_{q,y}\}$                 
        \item for all $q,r\in N(v)$ with $q\neq r$ and all $v_{q,x}\in L_q$ and $v_{r,y}\in L_r$, add the edge $\{v_{q,x}, \, v_{r,y}\}$
        \item for each $k\in N(v)$ and all $v_{x,y}\notin L_k$, add the edge $\{v_{x,y}, \, k\}$
        \item for all $x,y\in N(v)$, add the edge $\{x, \, y\}$ to extend $N(v)$ to a clique
    \end{itemize}}{$\aw(G) = \aw(G') + \w(v)$}{ If $\I'\cap N(v) = \emptyset$ then $\I=\I' \cup \{v\}$, else $\I = (\I'\cap V) \cup \{y \mid v_{x,y}\in \I'\setminus V\}$}
\end{reduction}
The Reduction~\ref{red:struction_modified}
is extended by removing the weight restriction for the vertex $v$ resulting in Reduction~\ref{red:struction_extended}.
\begin{reduction}[Extended Weighted Struction by Gellner~\etal~\cite{gellner2021boosting}]\phantomsection\label{red:struction_extended}
    Let $v\in V$ and $C$ the set of all independent sets $c$ in $G[N(v)]$ with $\w(v) < \w(c)$, \hbox{then transform $v$.} \\ 
\reductiondetails{Construct the graph $G'$ as follows
    \begin{itemize}
        \item remove $N[v]$ 
        \item for each $c \in C$
        \begin{itemize}
            \item add a vertex $v_c$ with $\w(v_c) = \w(c) - \w(v)$ 
            \item for each $w \in N(c) \setminus N(v)$ add the edge $\{w,\,v_c\}$ 
            \item for each $c'\in C\setminus\{c\}$, add the edge $\{v_c,\,v_{c'}\}$, forming a clique
        \end{itemize}
    \end{itemize}}{$\aw(G) = \aw(G') + \w(v)$}{
    If $\I'\cap V = \{v_c\}$, then 
    $\I = (\I'\cap V) \cup c$, else $\I = \I' \cup \{v\}$.} 
\end{reduction}
To potentially reduce the number of vertices, the authors in~\cite{gellner2021boosting} also proposed Reduction~\ref{red:struction_extended_reduced}. In this rule, they restrict the independent sets $c\in C$ used in Reduction~\ref{red:struction_extended} with an additional weight constraint. With this additional restriction, for a vertex $v$ only independent sets "just" greater than the weight $\w(v)$ are used to create new vertices.
\begin{reduction}[Extended Reduced Weighted Struction by Gellner~\etal~\cite{gellner2021boosting}]\label{red:struction_extended_reduced}
Let $C$ be the set of all independent sets in $G[N(v)]$. We define the set $C' = \{c \in C \mid \nexists c'\in C \text{ such that } c' \subsetneq c \text{ and } \w(c') > \w(v) \}$ as the set of independent set with weight "just" greater than $\w(v)$, then transform $v$. \\ 
\reductiondetails{Construct the graph $G'$ as follows
    \begin{itemize}
        \item remove $N[v]$ 
    \item for each independent set $c \in C'$  
    \begin{itemize}
        \item add a vertex $v_c$ with weight $\w(v_c) = \w(c) - \w(v)$; call the set of these added vertices $V_{C'}$
        \item for each $y\in N(v)\setminus N(c)$, add a vertex $v_{c,y}$ with weight $\w(v_{c,y}) = \w(y)$; call the set of these added vertices $V_E$
        \item for each $w \in N(c) \setminus N(v)$ add the edge $\{w,\,v_c\}$ 
        \item for all ${v_{c,y}\in L_c}$ and ${w \in (N(y) \cup N(c)) \setminus N(v)}$ add the edge $\{w,\,v_{c,y}\}$ 
    \end{itemize}
    \item for each $v_c \in V_{C'}$ and all $v_{c,x},v_{c,y}\in L_c$, if $\{x,\,y\}\in E$, then add the edge~$\{v_{c,x}, \, v_{c,y}\}$      
    \item for each $v_{c, y} \in V_E$ and all $v_{c'} \in V_{C'} \setminus \{v_c\}$, add the edge $\{v_{c'},\,v_{c,y}\}$      
    \item for all $v_c,v_{c'} \in V_{C'}$, add the edge $\{v_c,\,v_{c'}\}$, such that $V_{C'}$ is forming a clique
    \item for all $v_{c,y},v_{c',y'} \in V_E$ with $c \neq c'$, add the edge~$\{v_{c,y},\,v_{c',y'}\}$ 
    \end{itemize}}{$\aw(G) = \aw(G') + \w(v)$}{
    If $\I'\cap V_{C'} = \emptyset$, then $\I = \I' \cup \{v\}$, else replace the one vertex $v_c\in\I'\cap V_{C'}$ with the vertices in $c$ and all vertices $v_{c,y}\in \I'\cap V_E$ with $y$ resulting in $\I = \I'\setminus (V_{C'}\cup V_E) \cup c \cup \{y \mid v_{c,y} \in \I'\cap V_E\}$}
\end{reduction}

\subsection{Global Rules}
\label{sec:global-rules}
The following data reductions are not local reductions, but could potentially extend to \hbox{the entire graph.}
\begin{reduction}[Simultaneous Set by Xiao~\etal~\cite{xiao2021efficient}]\phantomsection\label{red:simultaneous}
    A set of vertices $S\subseteq V$ such that there is an MWIS that either contains all or non of the vertices in $S$ is called \emph{simultaneous}.
    Let $S\subseteq V$ be a simultaneous set, then fold $S$ into a new vertex $v'$.\\
    \reductiondetails{$G'=G[(V\cup \{v'\})\setminus S]$ with $\w(v')=\w(S)$ and $N(v')=N(S)$}{$\aw(G) = \aw(G')$}{If $v'\in \I'$, then $\I = \I' \cup S$, else $\I = \I'$}    
\end{reduction}
As introduced above, Reduction~\ref{red:simultaneous} is a meta-reduction that is not used in practice. However, in the following, we cover new rules that are special cases of this reduction.

For the next rule, a vertex is assumed to be part of all MWIS. If this assumption leads to a contradiction, the vertex can be excluded following the described algorithm.
\begin{reduction}[Unconfined Vertices by Xiao~\etal~\cite{xiao2021efficient}]	\label{red:unconfined}%
    A vertex $v$ can be excluded if a contradiction is obtained from the assumption that every maximum weight independent set of $G$ includes $v$. Let $S$ be an independent set of $G$. A vertex $x \in N(S)$ is called a child of $S$ if $\w(x) \geq \w(S \cap N(x))$ and a child is called an extending child if $|N(x) \setminus N[S]|=1$. For an extending child, the only vertex $y \in N(x) \setminus N[S]$ is called a satellite of $S$. Starting with $S=\{v\}$, we can find a contradiction by repeatedly extend $S$ with a satellite until one of the following conditions hold:\newpage
    \begin{enumerate}
        \item There exists a child $x$ such that $N(x) \setminus N[S] = \emptyset$
        \item All children $x \in S$ have $|N(x) \setminus N[S]|>1$
    \end{enumerate}      
    In the second case, the set $S$ confines $v$, meaning every maximum weight independent set that contains $v$, also contains $S$ and we can not reduce.
    In the first case, $v$ is called unconfined and can be excluded.\\
        \reductiondetails{$G'=G-v$}{$\aw(G) = \aw(G')$}{$\I = \I'$}   
\end{reduction}

\begin{reduction}[Extended Unconfined by Gro{\ss}mann~\etal~\cite{grossmann2025accelerating}]\phantomsection\label{red:e_unconfined}
    A vertex $v \in V$ can be removed if it is unconfined proven by the following procedure.
    Start with a set $S=\{v\}$. We assume $S$ is contained in every MWIS in $G$. We can search for a contradiction by repeatedly extending $S$ with satellites from one extending child until one of the following conditions hold
    \begin{enumerate}
        \item \label{redcase:unconfined1} There exists a child $x$ such that \\ $\w(x) \geq \w(S \cap N(x)) + \aw(G[N(x) \setminus N[S]])$.
        \item \label{redcase:unconfined2} There exist no further satellites to extend $S$.
    \end{enumerate}
    In the second case, the set $S$ confines $v$, meaning every maximum weight independent set that contains $v$, also contains $S$ and we can not remove $v$.
    In the first case, $v$ is called unconfined and can be excluded.\\
    \reductiondetails{$G'=G-v$}{$\aw(G) = \aw(G')$}{$\I = \I'$}
\end{reduction}

After having computed the confining sets in Reduction~\ref{red:unconfined}, we can reduce the instance further by working with these sets. In~\cite{xiao2021efficient} Xiao~\etal~introduced the notion of a 
\emph{simultaneous} set, which is a set of vertices $\{u,\, v\}\subseteq V$, where $u$ is in the confining set of $v$ and $v$ is in the confining set of $u$, \ie $u \in S_v$ and $v \in S_u$. In this case, both vertices are either in all MWIS or can both be excluded. Therefore, these vertices can be folded, as described \hbox{in Reduction~\ref{red:simUnconfined}.}
\begin{reduction}[Simultaneous Confined by Xiao~\etal~\cite{xiao2021efficient}]\phantomsection\label{red:simUnconfined}
Let $u,v\in V$ and $S_u$, $S_v$ be their corresponding confining sets computed by Reduction~\ref{red:unconfined}. If $u\in S_v$ and $v\in S_u$, then fold $u$ and $v$ into a new vertex $v'$.\\
        \reductiondetails{ $G' = G[(V\cup \{v'\})\setminus \{u,\,v\}]$ with $\w(v') = \w(u) + \w(v)$ and $N(v')=N(\{u, v\})$}{$\aw(G) = \aw(G')$}{If $v'\in \I'$, then $\I = (\I'\setminus \{v'\})\cup \{u,\,v\}$, else $\I = \I'$}
\end{reduction}

The next rule works analogue to Reduction~\ref{red:unconfined}, however, here we assume a vertex to be part of \emph{no} MWIS. If a contradiction is found, we include the vertex.
\begin{reduction}[Uncovered Vertices by Liu~\etal~\cite{liu2023application}]\label{red:cover}%
    A vertex $v$ can be included if a contradiction is obtained from the assumption that no maximum weight independent set of $G$ includes $v$. Let $C\subset V$ be a set of vertices that are in no MWIS of $G$.
    For a vertex $x\in C$ we define a \emph{mirror} as a vertex $y \in N^2(x)$ satisfying 
    $\w(x) \geq \aw(G[N(x)\setminus (C \cup N(y))])$. 
    Starting with $C=\{v\}$, we can find a contradiction by repeatedly extending $C$ with mirrors until one of the following conditions hold:
    \begin{enumerate}
        \item There exists a vertex $y \in C$ such that $\w(y) \geq \aw(G[N(y)\setminus C])$
        \item There are no mirrors to extend the set $C$
    \end{enumerate}      
    In the second case, the set $C$ covers $v$, meaning every maximum weight independent set that does not contain $v$, also does not contain $C$ and we can not reduce.
    In the first case, $v$ is called uncovered and can be included.\\
        \reductiondetails{$G'=G-N[v]$}{$\aw(G) = \aw(G')+\w(v)$}{$\I = \I'\cup \{v\}$}   
\end{reduction}
\begin{reduction}[Simultaneous Cover by Liu~\etal~\cite{liu2023application}]\phantomsection\label{red:simCover}
Let $u,v\in V$ and $S_u$, $S_v$ be their corresponding covering sets computed by Reduction~\ref{red:cover}. If $u\in S_v$ and $v\in S_u$, then fold $u$ and $v$ into a new vertex $v'$.\\
        \reductiondetails{$G' = G[(V\cup \{v'\})\setminus \{u,\,v\}]$ with $\w(v') = \w(u) + \w(v)$ and $N(v')=N(\{u, v\})$}{$\aw(G) = \aw(G')$}{If $v'\in \I'$, then $\I = (\I'\setminus \{v'\})\cup \{u,\,v\}$, else $\I = \I'$}
\end{reduction}

The next two reduction rules are based on small vertex-cuts of a graph. If there is such a cut, one component of the graph can be solved for all solution combinations in the cut and folded accordingly into new vertices.
\begin{reduction}[One Vertex Cut by Xiao~\etal~\cite{xiao2024maximum}]\phantomsection\label{red:cut_vertex}
		Let $v$ be an articulation point in $G$ and $G^*$ a connected component in $G-v$. Let $\I_1$ be the optimal solution on $G^*$ and $\I_2$ the optimal solution on $G^*-N(v)$.\begin{itemize}
		\item If $\w(v) + \w (\I_2) \leq \w(\I_1)$, then exclude $v$ and include the vertices in $\I_1$.\\
        \reductiondetails{$G' = G-(V(G^*) \cup \{v\})$}{$\aw(G) = \aw(G') + \w(\I_1)$}{$\I = \I' \cup \I_1$ \\[-.5em]}
		\item Else, fold $v$ and $G^*$ to a new vertex $v'$.\\
        \reductiondetails{$G' = G[(V \cup \{v'\})\setminus V(G^*) ]$ with $\w(v') = \w(v)+\w(\I_2) - \w(\I_1)$}{$\aw(G) = \aw(G') + \w(\I_1)$}{
		If $v'\in \I'$, then $\I = (\I'\setminus\{v'\})\cup \{v\}\cup \I_2$, else $\I = \I'\cup \I_1$}
        \end{itemize}  
\end{reduction}
\begin{reduction}[Two Vertex Cut by Xiao~\etal~\cite{xiao2024maximum}]\phantomsection\label{red:cut_vertex_two}
		Let ${u,v}$ be a vertex cut, \ie after removing $u$ and $v$ the graph $G$ is disconnected into two components. Let $G^*$ be a connected component in $G-\{u,v\}$. Let the following sets be MWIS for the respective subgraphs, $\I_v$ for $G^*-N[v]$, $\I_u$ for $G^*-N[u]$, $\I_{u,v}$ for $G^*-N[\{u,v\}]$ and $\I^*$ for $G^*$.
        We assume w.l.o.g. that $\w(\I_u) \geq \w(\I_v)$, then fold $G^*$ into \hbox{new vertices $x_u, x_v, x_{u,v}$.} \\
        \reductiondetails{ Construct $G'$ by
        \begin{itemize}
            \item removing $G^*$
            \item adding $x_u$ with $\w(x_u) = \w(\I_u)-\w(\I_{u,v})$
            \item adding $x_v$ with $\w(x_v) = \w(\I_v)-\w(\I_{u,v})$
            \item adding $x_{u,v}$ with $\w(x_{u,v}) = \w(\I^*)-\w(\I_u)$
            \item adding edges $\{v, \, x_u\}$, $\{x_u, \, x_v\}$, $\{x_v, \, u\}$, $\{v, \, x_{u,v}\}$ and $\{u, \, x_{u,v}\}$
        \end{itemize}
        }{$\aw(G) = \aw(G') + \w(\I_{u,v})$}
        {If $u\notin \I'$ and $v \in \I'$, then $\I = (\I' \setminus \{x_v\}) \cup \I_v$, \\& else if $u\in \I'$ and $v\notin \I'$, then $\I = (\I' \setminus \{x_u\}) \cup \I_u$, \\& else if $u,v\in \I'$, then $\I = \I' \cup \I_{u,v}$, \\& else $\I=(\I' \cap V) \cup \I^*$}
\end{reduction}
\begin{remark}
    For the Reductions~\ref{red:cut_vertex} and~\ref{red:cut_vertex_two} the authors additionally impose a bound for the component $G^*$.
\end{remark}

\paragraph*{Critical Weight Independent Set}

The critical weight independent set is a costly but powerful reduction rule. It covers the related crown reductions for the vertex cover problem~\cite{chlebik2008crown, fellows2003blow}. It is also closely related to the known fact that an optimal solution to the LP-relaxation is always half-integral~\cite{nemhauser1974properties}, \ie the optimal solution will always have each vertex as 0, 1, or $\frac{1}{2}$. The vertices that are 1 or 0 in such an optimal LP solution can be included or excluded, respectively. The following critical set reduction describes this scenario using the notion of a critical weight IS.
\begin{reduction}[CWIS by Butenko and Turkhanov~\cite{butenko-trukhanov}]\phantomsection\label{red:CWIS}
	Let $\I_c\subseteq V$ be a critical weighted IS of $G$, \ie $\w(\I_c) - \w(N(\I_c)) = \max\{\w(\I)-\w(N(\I)) \mid \I \text{ is an IS of }G \}$, then include vertices in $\I_c$.\\ 
        \reductiondetails{$G'=G-N[\I_c]$}{$\aw(G) = \aw(G')+\w(\I_c)$}{$\I = \I'\cup \I_c$}  
\end{reduction}

\subfile{../tikz/reductions/cwis.tex}

Despite the short definition, it is probably the most complicated rule to implement out of all the rules presented in this survey. Ageev~\cite{ageev1994finding} introduced a polynomial time algorithm for how to find a critical weight IS. In the following, we give an outline of that algorithm. To start, consider the following ILP formulation. For this, we use two binary vectors $X$ and $Y$, where $X$ represents the vertices in $\I_c$ and $Y$ the vertices $N(\I_c)$.
Because $X$ and $Y$ are binary vectors representing these sets, in saying add $u$ to $X$ we mean set the $u$-th index in $X$ to 1. 
The elements in $X$ are all vertices where the corresponding index is set to 1.
\[\max \sum_{u \in V} \w(u) X_u - \w(u) Y_u \]
\[\text{s.t.} \quad Y_u \geq X_v \qquad \forall \, \{u, v\} \in E \]
\[X_u, Y_u \in \{0, 1\} \qquad \forall \, u \in V \]
In this formulation, for each vertex $u$ added to $X$, the neighborhood $N(u)$ must be added to $Y$. 
The objective value for an optimal solution to this ILP is non-negative since adding all vertices to $X$ and $Y$ is a feasible solution of weight zero. 
Note that with this definition an optimal solution is not guaranteed to be an independent set. However, as Ageev points out, we can always find an independent set with the exact same objective value by selecting all isolated vertices in the induced graph obtained from the elements in $X$. This is true because any vertices in $X$ with other neighbors in $X$ must also be in $Y$. Therefore, these vertices contribute exactly zero to the objective value. We introduce this ILP formulation because it is an instance of a simpler problem called the \textsc{Selection} problem that can be solved in polynomial time~\cite{balinski1970selection}.

Balinski introduced an algorithm for the \textsc{Selection} problem that we can use to identify a CWIS directly~\cite{balinski1970selection}. The algorithm solves a \textsc{Maximum Flow} problem on a special flow graph, constructed from the original graph. An illustration of this algorithm is shown in Figure~\ref{fig:cwis}. The flow graph construction starts with a directed bipartite graph with two sets of the original vertices $V$ and $V'$. For each edge $\{u, v\} \in E$ in the original graph, we add a directed edge from $u \in V$ to $v \in V'$ with infinite capacity in the bipartite graph. Then, add two more vertices $s$ and $t$. For each vertex $u \in V$, add the directed edge $(s, u)$ with capacity $\w(u)$, and for each vertex $u \in V'$, add the directed edge $(u, t)$ with capacity $\w(u)$. In this flow graph, we want to find a minimum cut. By removing the edges in this minimum cut, any vertices in $V$ that can still be reached from $s$ make up our critical weight IS. As shown in Figure~\ref{fig:cwis}, we can do this directly by running a BFS or DFS from $s$ in the residual graph after solving the \textsc{Maximum Flow} problem.

\subsection{Twin Based Reduction Rules}
\label{sec:further-reduction-rules}
Two vertices $u$ and $v$ which are not connected but have the same neighborhood are called twins. These present a special case of Reduction~\ref{red:simultaneous} and can be reduced by the following reduction. With the use of a hash function this reduction can be checked very efficiently.
\begin{reduction}[Twin by Lamm~\etal~\cite{lamm2019exactly}]\phantomsection\label{red:twin}
	Let $u, v\in V$ have equal, independent neighborhoods $N(u) = N(v) = \{p,\,q,\,r\}$.
	\begin{itemize}
		\item If $\w(\{u,\,v\}) \geq \w(\{p,\,q,\,r\})$, then include $u$ and $v$.\\
        \reductiondetails{$G' = G[V\setminus N[\{u,v\}]$}{$\aw(G)=\aw(G') +\w(u) + \w(v)$}{$\I = \I' \cup \{u,\,v\}$ \\[-.5em]}
		\item If $\w(\{u,v\}) < \w(\{p,q,r\})$ but $\w(\{u,v\}) > \w(\{p,\,q,\,r\}) - \min\{\w(x) \mid x\in\{p,\,q,\,r\}\}$, then fold $u,\, v,\, p,\, q,\, r$ into a new vertex $v'$. \\ 
        \reductiondetails{ $G' = G[(V\cup \{v'\})\setminus(N[v]\cup \{u\})]$ with $\w(v') = \w(\{p,\,q,\,r\}) - \w(\{u,\,v\})$ and $N(v')=N(\{p,\,q,\,r\})$}{$\aw(G) = \aw(G') + \w(\{u,\,v\})$}{If $v'\in \I'$, then $\I = (\I'\setminus \{v'\})\cup \{p,\,q,\,r\}$, else $\I = \I' \cup \{u,\,v\}$ \\[-.5em]}
	\end{itemize}
\end{reduction}
    Reduction~\ref{red:twin}, which is a special case of Reduction~\ref{red:simultaneous}, can be also applied if the neighborhood is larger~\cite{lamm2019exactly}.


\begin{reduction}[Extended Twin]\phantomsection\label{red:e_twin}
    Let $u, v\in V$ be non-adjacent and with equal neighborhoods $N(u) = N(v)$. Let further $\I_{N(v)}$ be an MWIS on $G[N(v)]$.
    \begin{itemize}
        \item If $\w(u)+\w(v) \geq \w(\I_{N(v)})$, then include $u$ and $v$.\\
              \reductiondetails{$G' = G-N[\{u,v\}]$}{$\aw(G) = \aw(G') + \w(u) + \w(v)$}{$\I = \I' \cup \{u, v\}$\\[-.5em]}
        \item If $\w(u) + \w(v) < \w(\I_{N(v)})$ but $\I_{N(v)}$ is the only independent set in $N(v)$ with this property, then fold $u$, $v$, and $N(v)$ into $v'$.\\
              \reductiondetails{${G' = G[(V \cup \{v'\})\setminus N[\{u,v\}]]}$ with ${N(v') = N(N[\{u,v\}])}$ and \\ &${\w(v') = \w(\I_{N(v)}) - \w(u) - \w(v)}$}{$\aw(G) = \aw(G') + \w(u) + \w(v)$}{If $v' \in \I'$, then $\I = (\I' \setminus \{v'\})\cup \I_{N(v)}$, else $\I = \I' \cup \{u, v\}$\\[-.5em]}
        \item Otherwise, fold $u$ and $v$ into a new vertex $v'$.\\
              \reductiondetails{$G' = G[(V\cup \{v'\}) \setminus \{u,v\} ]$ with $N(v') = N(\{u,\,v\})$ and $\w(v') = \w(v) + \w(u)$}{$\aw(G) = \aw(G')$}{If $v' \in \I'$, then $\I = \I' \setminus \{v'\} \cup \{u,\,v\}$, else $\I = \I'$}
    \end{itemize}
\end{reduction}

\begin{figreduction}[Almost Twin]{a_twin}
    Let $u, v \in V$ be non-adjacent such that $N(u) \subseteq N(v)$. If $\w(u) + \w(v) \geq \w(N(v))$, \hbox{then include $u$.}\\
    \reductiondetails{$G' = G - N[u]$}{$\aw(G) = \aw(G') + \w(u)$}{$\I = \I' \cup \{u\}$}
\end{figreduction}

\begin{figure}[t]
    \centering
    	\begin{tikzpicture}
		\node[rnode] (z) at (5,4) {};
		\node[rnode=$v$] (v) [right=0.75cm of z] {};
		\node[rnode] (x) [below=0.6cm of z] {};
		\node[rnode] (y) [above=0.6cm of z] {};
		\node[rnode=$w$] (w) [below=0.6cm of v] {};
		\node[lnode=$u$] (u) [left=.75cm of z] {};
		\remainingG{5,2}{$G-(N[v]\cup \{u\})$}
		\draw[edge] (w) -- (5.95,2);
		\draw[edge] (w) -- (5.5,2);
		\draw[edge] (x) -- (5.5,2);
		\draw[edge] (x) -- (5.175,2);
		\draw[edge] (x) -- (4.85,2);
		\draw[edge] (w) -- (v);
		\draw[edge] (v) -- (x);
		\draw[edge] (v) -- (y);
		\draw[edge] (v) -- (z);
		\draw[edge] (u) -- (x);
		\draw[edge] (u) -- (y);
		\draw[edge] (u) -- (z);
		\draw[edge] (y) -- (z);
		\draw[edge] (x) -- (4.4,2);
		\node[weight] (vweight) at ($(v) + (.2, -.1)$) {$4$};
		\node[weight] (uweight) at ($(u) + (-.2, -.1)$) {$8$};
  		\node[weight] (xweight) at ($(x) + (0, -.1)$) {$2$};
  		\node[weight] (yweight) at ($(y) + (0, -.1)$) {$4$};
  		\node[weight] (zweight) at ($(z) + (0.25, -.1)$) {$3$};
  		\node[weight] (wweight) at ($(w) + (.25, -.1)$) {$2$};
		\draw[black,-{Stealth[scale=1.5]}] ($(v) + (0.75,0)$) -- ($(v) + (1.75,0)$) ;	
	    \begin{scope}[xshift=4cm]
     
		\node[nodeE] (z) at (5,4) {};
		\node[rnode=$v$] (v) [right=0.75cm of z] {};
		\node[nodeE] (x) [below=0.55cm of z] {};
		\node[nodeE] (y) [above=0.6cm of z] {};
		\node[rnode=$w$] (w) [below=0.6cm of v] {};
		\node[nodeI] (u) [left=.75cm of z] {};
		\remainingG{5,2}{$G-(N[v]\cup \{u\})$}
		\draw[edge] (w) -- (5.95,2);
		\draw[edge] (w) -- (5.5,2);
		\draw[edgeR] (x) -- (5.5,2);
		\draw[edgeR] (x) -- (5.175,2);
		\draw[edgeR] (x) -- (4.85,2);
		\draw[edge] (w) -- (v);
		\draw[edgeR] (v) -- (x);
		\draw[edgeR] (v) -- (y);
		\draw[edgeR] (v) -- (z);
		\draw[edgeR] (u) -- (x);
		\draw[edgeR] (u) -- (y);
		\draw[edgeR] (u) -- (z);
		\draw[edgeR] (y) -- (z);
		\draw[edgeR] (x) -- (4.4,2);
		\node[weight] (vweight) at ($(v) + (.2, -.1)$) {$4$};
  		\node[weight] (wweight) at ($(w) + (.25, -.1)$) {$2$};
	    \end{scope}
	\end{tikzpicture}
    \caption{Illustration  from~\cite{grossmann2025accelerating} showing the Almost Twin (Reduction~\ref{red:a_twin}). In the original graph on the left, $N(u) \subseteq N(v)$ with $\w(u) + \w(v) \geq \w(N(v))$. By applying Reduction~\ref{red:a_twin}, $u$ is included and its neighbors excluded from $\I$, resulting in the reduced graph on the right.
    \vspace{-1em}}\label{fig:a_twin}
\end{figure}

\section{Discussion of Data Reductions in Practice}

This section focuses on the practical application of data reductions for the MWIS and MWVC problems. First, we discuss the relations between the reductions and their computational cost in practice. Then, we survey the use of these data reduction rules in practical solvers developed for these problems.

\begin{table}[!th]
\caption{Overview of all data reduction rules grouped by their type. We give additional information about where they are (first) introduced and on what page of the paper. Reduction rules marked with \emph{code} were not formally introduced in a paper, but were implemented in the associated source code.}
  \label{tab:rules_overview}
  \vspace{-0.5em}
  \begin{tabular}{lm{5.cm}lm{2.6cm}m{.7cm}m{.5cm}}
    \textbf{Ref.} & \textbf{Reduction Name} & \textbf{Type} & \textbf{Introduced By} & \textbf{In} & \textbf{At} \\ \midrule
    \ref{red:deg1} & Degree One & Low Degree & Gu~\etal~\cite{gu2021towards} & 2021 & p.4 \\
    \ref{red:triangle} & Triangle & Low Degree& Gu~\etal~\cite{gu2021towards} & 2021 & p.5\\    
    \ref{red:vShape} & V-Shape & Low Degree& Lamm~\etal~\cite{lamm2019exactly} \hfill \break and Gu~\etal~\cite{gu2021towards} & 2019 2021 & p.7\hspace{.5cm} p.5\\    
    \ref{red:path} & 3-Path & Low Degree & Xiao~\etal~\cite{xiao2024maximum}& 2024 & p.10\\
    \ref{red:4path} & 4-Path & Low Degree& Xiao~\etal~\cite{xiao2024maximum}& 2024 & p.12\\
    \ref{red:4cycle} & 4-Cycle & Low Degree& Xiao~\etal~\cite{xiao2024maximum}& 2024 & p.11\\
    \ref{red:5cycle} & 5-Cycle & Low Degree& Xiao~\etal~\cite{xiao2024maximum}& 2024 & p.13\\
    \ref{red:6cycle} & 6-Cycle & Low Degree& Xiao~\etal~\cite{xiao2024maximum}& 2024 & p.15\\ [0.5em]
    \ref{red:neighborhood_meta} & Heavy Vertex & Neighborhood& Lamm~\etal~\cite{lamm2019exactly} & 2019 & p.5\\       
    \ref{red:neighborhood removal} & Neighborhood Removal & Neighborhood& Lamm~\etal~\cite{lamm2019exactly} & 2019 & p.6\\    
    \ref{red:clique_neighborhood} & Clique Neighborhood Removal & Neighborhood& Lamm~\etal~\cite{kamis,lamm2019exactly} & 2019 & code\\    
    \ref{red:neighborhood folding} & Neighborhood Folding & Neighborhood & Lamm~\etal~\cite{lamm2019exactly} & 2019 & p.5\\      
    \ref{red:generalized_fold} & Generalized Neighborhood Folding & Neighborhood& Lamm~\etal~\cite{kamis,lamm2019exactly} & 2019 & code\\       
    \ref{red:2vertex neighborhood removal} & Two Vertex Neighborhood Removal & Neighborhood & Zheng~\etal~\cite{zheng2020efficient}& 2020 & p.3\\    
    \ref{red:heavy_set} & Heavy Set & Neighborhood & Xiao~\etal~\cite{xiao2021efficient} &2021 & p.4 \\
    \ref{red:heavy_set3} & Extended Heavy Set & Neighborhood & Gro{\ss}mann~\etal~\cite{grossmann2025accelerating} &2025 & p.13 \\ [0.5em]
    \ref{red:clique} & Simplicial Vertex & Clique &  Lamm~\etal~\cite{lamm2019exactly} & 2019 & p.6\\     
    \ref{red:clique_weight_transfer} & Simplicial Weight Transfer & Clique & Lamm~\etal~\cite{lamm2019exactly} & 2019 & p.6\\ 
    \ref{red:wFunnel} & Weighted Funnel & Clique & Gro{\ss}mann~\etal~\cite{grossmann2025accelerating} & 2025 & p.12\\  [0.5em]
    \ref{red:e_dom_rev} & Extended Domination Reversed & Domination & Gro{\ss}mann~\etal~\cite{grossmann2025accelerating} & 2025 & p.9\\    
    \ref{red:dom} & Domination & Domination & Lamm~\etal~\cite{lamm2019exactly} & 2019 & p.7\\    
    \ref{red:bse} & Basic Single Edge & Domination & Gu~\etal~\cite{gu2021towards} &2021 & p.6\\    
    \ref{red:ese} & Extended Single Edge & Domination & Gu~\etal~\cite{gu2021towards} & 2021 & p.6\\     [0.5em]
    \ref{red:e_dom} & Extended Domination & Domination & Gro{\ss}mann~\etal~\cite{grossmann2025accelerating} & 2025 & p.8\\    
    \ref{red:e_dom_rev} & Extended Domination Reversed & Domination & Gro{\ss}mann~\etal~\cite{grossmann2025accelerating} & 2025 & p.9\\    
    \ref{red:struction_original} & Original Weighted Struction & Struction & Gellner~\etal~\cite{gellner2021boosting} &2021 & p.4 \\   
    \ref{red:struction_modified} & Modified Weighted Struction & Struction & Gellner~\etal~\cite{gellner2021boosting} &2021 & p.4\\    
    \ref{red:struction_extended} & Extended Weighted Struction & Struction & Gellner~\etal~\cite{gellner2021boosting} &2021 & p.5\\   
    \ref{red:struction_extended_reduced} & Extended Reduced Weighted \hfill \break Struction & Struction & Gellner~\etal~\cite{gellner2021boosting} & 2021 & p.5 \\    [0.5em] 
    \ref{red:simultaneous} & Simultaneous Set & Global & Xiao~\etal~\cite{xiao2021efficient} & 2021 & p.4  \\    
    \ref{red:unconfined} & Unconfined Vertices & Global &  Xiao~\etal~\cite{xiao2021efficient} & 2021 & p.4 \\  
    \ref{red:e_unconfined} & Extended Unconfined Vertices & Global &  Gro{\ss}mann~\etal~\cite{grossmann2025accelerating} & 2025 & p.14 \\  
    \ref{red:simUnconfined} & Simultaneous Confined & Global &  Xiao~\etal~\cite{xiao2021efficient} & 2021 & p.4 \\
    \ref{red:cover} & Uncovered Vertices & Global & Liu~\etal~\cite{liu2023application} & 2023 & p.16 \\ [0.5em]
    \ref{red:simCover} & Simultaneous Cover & Global &Liu~\etal~\cite{liu2023application} & 2023  & p.17 \\    
    \ref{red:cut_vertex} & One Vertex Cut & Global & Xiao~\etal~\cite{xiao2024maximum} & 2024 & p.17 \\    
    \ref{red:cut_vertex_two} & Two Vertex Cut & Global & Xiao~\etal~\cite{xiao2024maximum} & 2024 & p.19 \\    
    \ref{red:CWIS} & Critical Weight Independent Set & Global & Butenko and \hfill \break Turkhanov~\cite{butenko-trukhanov} & 2007 & p.2 \\   [0.5em]  
    \ref{red:twin} & Twin & Twin & Lamm~\etal~\cite{lamm2019exactly} & 2019 & p.7\\ 
    \ref{red:e_twin} & Extended Twin & Twin & Gro{\ss}mann~\etal~\cite{grossmann2025accelerating} & 2025 & p.9\\ 
    \ref{red:a_twin} & Almost Twin & Twin & Gro{\ss}mann~\etal~\cite{grossmann2025accelerating} & 2025 & p.11
  \end{tabular}
  \vspace{-2em}
\end{table}


\begin{figure}[!ht]
    \centering
    \resizebox{\textwidth}{!}{
        \begin{tikzpicture}[scale=0.85,yscale=1.2,every node/.style={rectangle, align=center, draw=none, fill=lightgray, rounded corners=1.5mm}]

            \begin{scope}[every node/.style={rectangle, align=center, thick, draw=lipicsYellow, fill=lipicsYellow!10}]
                \node (simultan) at (-6.5, 2.5) {Simultaneous \\Set$^{\star}$ (\ref{red:simultaneous})};
            \end{scope}
            \begin{scope}[every node/.style={rectangle, align=center, draw=lipicsYellow, thick, fill=lipicsYellow, rounded corners=1.5mm}]
                \node (extended_unconfined) at (-2.5, -1.5) {Extended \\Unconfined (\ref{red:e_unconfined})};
                \node (cover) at (-8.5, -2.25) {Uncovered \\ Vertices (\ref{red:cover})};
                \node (2cut) at (-11.75, -2.25) {Two Vertex\\Cut (\ref{red:cut_vertex_two})};
                \node (1cut) at (-11.75, -4.5) {One Vertex\\Cut (\ref{red:cut_vertex})};
                \node (gen_fold) at (9.25, -2.25) {Generalized\\ Fold (\ref{red:generalized_fold})};
                \node (heavy_set) at (3.75, -2.25) {Heavy Set (\ref{red:heavy_set})};
                \node (heavy_set3) at (3.75, -1.) {Extended \\Heavy Set (\ref{red:heavy_set3})};
                \node (heavy_vertex) at (3.75, -4.5) {Heavy\\ Vertex (\ref{red:neighborhood_meta})};
                \node (scover) at (-8.5, -0.2) {Simultaneous \\Cover (\ref{red:simCover})};
                \node (e_twin) at (-6.5, -4.5) {Extended \\Twin (\ref{red:e_twin})};
            \end{scope}

            \begin{scope}[every node/.style={rectangle, align=center, draw=none, fill=lightgray!70, rounded corners=1.5mm}]
                \node (sconfined) at (-4.5, -6.5) {Simultaneous \\Confined (\ref{red:simUnconfined})};
                \node (critical_set) at (0, 0.) {Critical Set (\ref{red:CWIS})};
                \node (unconfined) at (-2.5, -8.25) {Unconfined \\ Vertices (\ref{red:unconfined})};
                \node (single_edge) at (-2.5, -11.5) {Basic Single Edge (\ref{red:bse})};
                \node (e_single_edge) at (-11, -11) {Extended \\Single Edge (\ref{red:ese})};
                \node (dom) at (-2.5, -12.5) {Domination (\ref{red:dom})};
                \node (e_dom) at (-10.5, -12.5) {Extended Domination \\ (Reversed) (\ref{red:e_dom}, \ref{red:e_dom_rev})};
                \node (2neighbor) at (0, -14.5) {Two Vertex \\Neighbor\\ Removal (\ref{red:2vertex neighborhood removal})};
                \node (neighbor) at (3.75, -14.25) {Neighbor \\Removal (\ref{red:2vertex neighborhood removal})};
                \node (clique_neigh) at (3.75, -9) {Clique \\Neighborhood \\Removal (\ref{red:clique_neighborhood})};
                \node (clique) at (5.75, -12) {Isolated \\Vertex (\ref{red:clique})};
                \node (clique_transfer) at (9.25, -12) {Isolated\\ Weight \\Transfer (\ref{red:clique_weight_transfer})};
                \node (funnel) at (12.45, -9) {Funnel (\ref{red:wFunnel})};
                \node (deg1) at (5, -17) {Degree One (\ref{red:deg1})};
                \node (triangle) at (9, -17) {Triangle (\ref{red:triangle})};
                \node (vShape) at (12.45, -17) {V-Shape (\ref{red:vShape})};

                \node (a_twin) at (-4.5, -10.5) {Almost\\ Twin (\ref{red:a_twin})};
                \node (twin) at (-6.5, -12.5) {Twin (\ref{red:twin})};

                \path[Stealth-] (scover) edge[out=90, in=270] (simultan);
                \path[Stealth-] (sconfined) edge[out=90, in=270] (simultan);
                \path[Stealth-] (e_twin) edge[out=90, in=270] (simultan);
                \path[Stealth-] (twin) edge[out=90, in=270] (e_twin);
                \path[Stealth-] (twin) edge[out=90, in=180, looseness=0.9] (a_twin.west);
                \path[Stealth-, dashed] (cover) edge[out=90, in=270] (scover);
                \path[Stealth-] (1cut) edge[out=90, in=270] (2cut);

                \path[Stealth-, dashed] (unconfined.40) edge[out=90, in=0] (sconfined.east);
                \path[Stealth-] (unconfined) edge[out=90, in=270] (extended_unconfined);
                \path[Stealth-] (single_edge) edge[out=90, in=270] (unconfined);
                \path[Stealth-] (dom) edge[out=90, in=270] (single_edge);

                \path[Stealth-] (2neighbor) edge[out=90, in=270] (critical_set);

                \path[Stealth-] (heavy_set) edge[out=90, in=270] (heavy_set3);
                \path[Stealth-] (heavy_vertex) edge[out=90, in=270] (heavy_set);
                \node[below = 6 of critical_set, fill=none] (p1) {};
                \path[-] (p1) edge[out=90, in=180,looseness=0.8] (heavy_set.south west);
                \path[-] (p1) edge[out=90, in=0] (cover.south east);

                \path[Stealth-] (clique_neigh) edge[out=90, in=270] (heavy_vertex);
                \path[Stealth-] (clique) edge[out=90, in=270] (clique_neigh);
                \path[Stealth-] (neighbor.north) edge[out=90, in=270] (clique_neigh.south);

                \path[Stealth-] (clique_transfer) edge[out=90, in=270] (gen_fold);
                \path[Stealth-] (funnel) edge[out=90, in=270] (gen_fold);

                \path[Stealth-] (vShape) edge[out=90, in=0] (neighbor.east);
                \path[Stealth-] (deg1) edge[out=90, in=270] (neighbor);

                \path[Stealth-] (triangle) edge[out=90, in=270] (clique_transfer.south);
                \path[Stealth-] (triangle) edge[out=90, in=270] (clique.south);
                \path[Stealth-] (deg1) edge[out=90, in=270] (clique_transfer.south);
                \path[Stealth-] (vShape) edge[out=90, in=270] (gen_fold.south);

            \end{scope}

        \end{tikzpicture}}
        \caption{This figure gives an overview of the presented reduction rules and their relations. Two rules A and B are connected with an arrow from A $\rightarrow$ B if rule A is a more generalized form and can also reduce the patterns reduced by Reduction B. A dashed arrow indicates that the rule B is part of rule A. The rules are intuitively sorted by complexity, starting with the more computationally expensive rules at the top. Hence, more general rules are always above their special case rules. Note that even though the critical set rule can be applied in polynomial time, it has to be applied to the whole graph. That makes the rule more computationally expensive than other rules that have to \noindent\colorbox{lipicsYellow}{solve the MWIS on a bounded subgraph,} marked in yellow. The Simultaneous Set rule is only added as a meta reduction and not implemented for the \hbox{most general case.} The struction-based rules are omitted since they transform the graph, so they are not easily comparable to the other reduction rules.}
        \label{fig:reduction_relation}
    \end{figure}
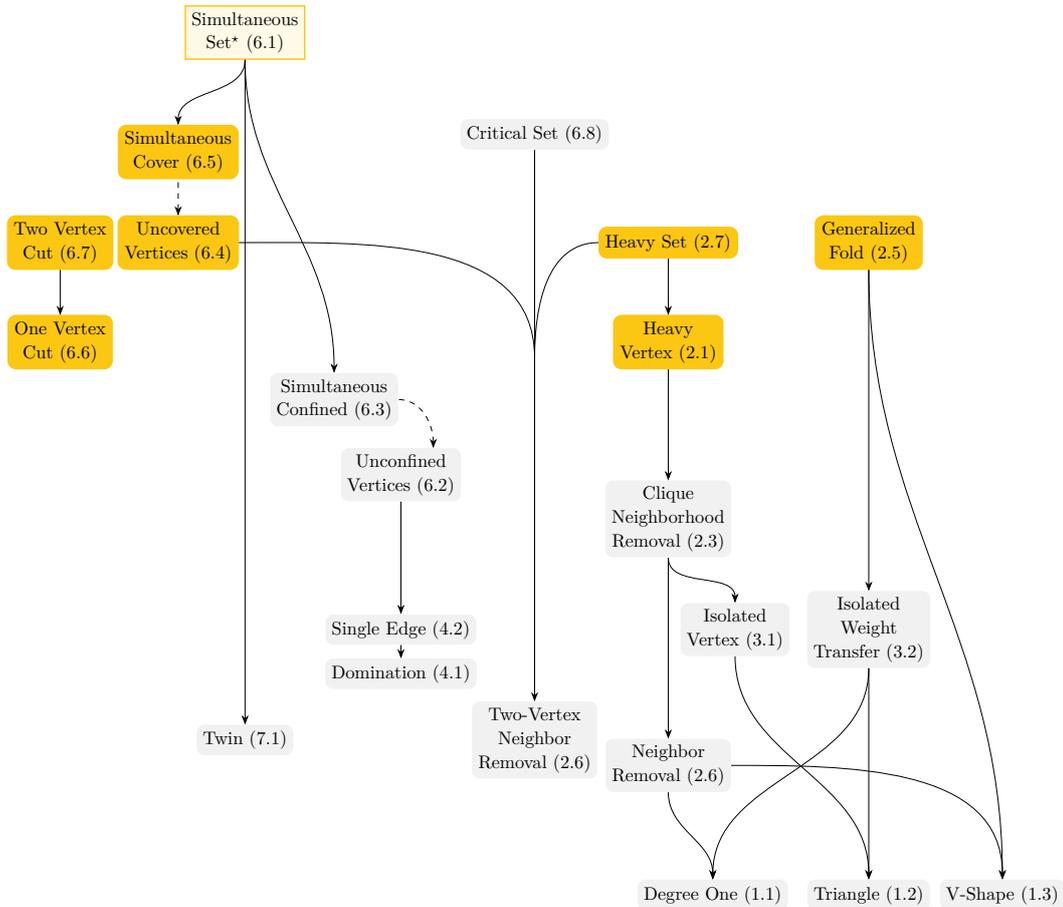

\subsection{Relations Between Data Reductions}
\label{sec:relation}
In the previous sections, we gave an overview of several different reduction rules with varying complexities, 
summarized in Table~\ref{tab:rules_overview}.
Some of these rules are fast, \eg Reduction~\ref{red:deg1}, while others, if not bound, have exponential running time (Reduction~\ref{red:cut_vertex}). Furthermore, most of the reduction rules are special cases of other, more general reduction rules.
This section discusses the relations between different reduction rules and gives a rough overview of their (practical) running times and complexities.
Figure~\ref{fig:reduction_relation} presents most of the introduced reductions and their relations. The reduction rules are approximately sorted by decreasing practical running times from top to bottom. 
The most general rule for simultaneous sets is ranked highest and only added as a meta reduction since there is no efficient way of finding general simultaneous sets.
The CWIS reduction has a polynomial running time but must be applied to the whole graph. Compared to other reductions that can be bounded and applied locally only for small neighborhoods, the CWIS reduction is more computationally expensive in practice, even if some of these other reductions have exponential running time if left unbounded (marked in yellow). Their performance heavily depends on the size bound for the subproblem to solve.
Among these yellow-marked rules, we sorted them according to how often an independent set has to be solved on the bounded subgraph. For example, the heavy set and generalized fold reductions have to solve multiple MWIS and are therefore considered slower than the heavy vertex reduction.
The Degree One, V-Shape, and Triangle are the fastest reduction rules. Note that all path and cycle rules in Section~\ref{sec:deg-bounded-rules} are covered by Triangle and V-Shape, but not necessarily faster and therefore omitted in this figure.

The arrows in Figure~\ref{fig:reduction_relation} describe the relation between different reduction rules. These are always directed from a \emph{general} to a \emph{special} case. Note the use of transitivity for this relation; therefore, some edges are omitted. For example, the Heavy Vertex rule covers the Clique Neighborhood Removal rule, which again covers the simpler Neighborhood Removal rule. Because of this, Neighborhood Removal is also a special case of Heavy Vertex. However, this transitivity does not apply to dashed arrows. For example, the Uncovered Vertices rule is used to compute the covering sets in Simultaneous Cover, a special case of Simultaneous Set. However, the more general Simultaneous Set rule can not necessarily reduce the patterns that the Uncovered Vertices rule does.
Furthermore, if a reduction rule has multiple cases, we add an arrow if only one is covered. For example, we need the Generalized Fold and the Neighborhood Removal rules to cover the V-Shape reduction fully.
Looking at the relations between the different rules, we see that the Heavy Set and Generalized Fold reductions are very powerful and cover all the low degree, clique, and neighborhood-based reduction rules. 
Then, there are reductions derived from the simultaneous set meta reduction, cut reductions, and the reductions Unconfined Vertices and Uncovered Vertices, which are not special cases \hbox{of other rules.}

\subsection{Practical Use of Data Reductions in Different Solvers}
\label{sec:use}

\newcommand{\cmark}{\ding{51}}
\begin{table}[t]
  \caption{Groups of data reduction rules used by different algorithms with implementations and experimental evaluation, sorted by year when the solver was first published. The reductions combined in the different groups are mentioned in the parenthesis after \hbox{the group name.}}
  \vspace{-1em}
  \label{tab:rules}
  \begin{FlushLeft}
    \begin{tabular}{l*{11}{ll}}
      \textbf{Year}                                                                                                                                         & \textbf{Method}                            &
      \multicolumn{1}{p{0.35cm}}{\rotatebox[origin=l]{45}{\textbf{Degree One (\ref{red:deg1})}}}                                                            &
      \multicolumn{1}{p{0.35cm}}{\rotatebox[origin=l]{45}{\textbf{Degree Two (\ref{red:triangle}-\ref{red:6cycle})}} }                                      &
      \multicolumn{1}{p{0.35cm}}{\rotatebox[origin=l]{45}{\textbf{Neighborhood (\ref{red:neighborhood removal} -\ref{red:2vertex neighborhood removal})} }} &
      \multicolumn{1}{p{0.35cm}}{\rotatebox[origin=l]{45}{\textbf{Domination (\ref{red:dom},\ref{red:bse},\ref{red:ese})}} }                                &
      \multicolumn{1}{p{0.35cm}}{\rotatebox[origin=l]{45}{\textbf{Clique (\ref{red:clique}-\ref{red:wFunnel})}} }                                           &
      \multicolumn{1}{p{0.35cm}}{\rotatebox[origin=l]{45}{\textbf{Twin (\ref{red:twin}-\ref{red:a_twin})}}}                                                 &
      \multicolumn{1}{p{0.35cm}}{\rotatebox[origin=l]{45}{\textbf{Unconfined (\ref{red:unconfined}-\ref{red:simCover})}} }                                  &
      \multicolumn{1}{p{0.35cm}}{\rotatebox[origin=l]{45}{\textbf{CWIS (\ref{red:CWIS})}}}                                                                  &
      \multicolumn{1}{p{0.35cm}}{\rotatebox[origin=l]{45}{\textbf{Struction (\ref{red:struction_original}-\ref{red:struction_extended_reduced})}}}          &
      \multicolumn{1}{p{0.35cm}}{\rotatebox[origin=l]{45}{\textbf{Heavy Set (\ref{red:heavy_set})}} }                                                       &
      \multicolumn{1}{p{0.35cm}}{\rotatebox[origin=l]{45}{\textbf{Cut (\ref{red:cut_vertex},\ref{red:cut_vertex_two})}}}
      \\ \midrule
      2018                                                                                                                                                  & {\numwvc{}}~\cite{li2020numwvc}            & \cmark & \cmark &        &        &        &        &        &        &        &        &        \\ 
      2019                                                                                                                                                  & {\kamis{}}~\cite{lamm2019exactly}          & \cmark & \cmark & \cmark & \cmark & \cmark & \cmark &        & \cmark &        &        &        \\ 
      2019                                                                                                                                                  & {\bmwvc{}}~\cite{wang2019exact}            & \cmark & \cmark & \cmark &        &        &        &        &        &        &        &        \\ 
      2020                                                                                                                                                  & {\maehts}~\cite{wang2021fast}              & \cmark &        & \cmark &        &        &        &        &        &        &        &        \\ 
      2020                                                                                                                                                  & {\dttwo}~\cite{zheng2020efficient}         &        &        & \cmark &        &        &        &        &        &        &        &        \\ 
      2021                                                                                                                                                  & {\solve}~\cite{xiao2021efficient}          &        & \cmark & \cmark &        &        & \cmark & \cmark & \cmark &        & \cmark &        \\ 
      2021                                                                                                                                                  & {\htwis}~\cite{gu2021towards}              & \cmark & \cmark &        & \cmark &        &        &        &        &        &        &        \\ 
      2021                                                                                                                                                  & {\struction}~\cite{gellner2021boosting}    & \cmark & \cmark & \cmark & \cmark & \cmark & \cmark &        & \cmark & \cmark &        &        \\ 
      2022                                                                                                                                                  & {\gnnvc}~\cite{langedal2022efficient}      & \cmark & \cmark & \cmark & \cmark & \cmark & \cmark &        & \cmark &        &        &        \\ 
      2023                                                                                                                                                  & {\mmwis}~\cite{grossmann2023finding}       & \cmark & \cmark & \cmark & \cmark & \cmark & \cmark &        & \cmark & \cmark & \cmark &        \\ 
      2024                                                                                                                                                  & {\cbnr/\csearch}~\cite{liu2023application} &        &        &        &        &        &        & \cmark &        &        &        &        \\ 
      2024                                                                                                                                                  & {\hglv}~\cite{tan2024efficient}            & \cmark &        & \cmark &        &        &        &        &        &        &        &        \\ 
      2025                                                                                                                                                  & {\dynls}~\cite{zhu2025dynamic}             & \cmark & \cmark & \cmark & \cmark & \cmark & \cmark &        & \cmark & \cmark &        &        \\ 
      2025                                                                                                                                                  & {\lnr}~\cite{grossmann2025accelerating}    & \cmark & \cmark & \cmark & \cmark & \cmark & \cmark & \cmark & \cmark & \cmark & \cmark & \cmark \\[-1.5em] 
    \end{tabular}
  \end{FlushLeft}
\end{table}

We now examine which data reduction rules are utilized and evaluated in practical implementations. In Table~\ref{tab:rules}, we list all the algorithms that use data reduction rules for solving the \textsc{Maximum Weight Independent Set} or \textsc{Minimum Weight Vertex Cover} problems in practice. For each solver, we mark which rules are utilized.
The most commonly used reduction rules are the degree bounded rules (Degree One, V-Shape, and Triangle) and the Neighborhood Removal rule. These rules are fast to compute and effective in practice. 
After these simple rules, domination-based reductions are applied most often. These reduction rules are used in 5 of the 12 algorithms considered.
The clique-based reductions and the Critical Weight Independent Set reductions are used in four different solvers.

The reduction rules Struction, Unconfined, and Heavy Set are used only in two algorithms. The cut-based rules, introduced in a theoretical paper~\cite{xiao2024maximum}, are not used in any practical solver. An explanation for why only a few implementations use these rules could be that they are more computationally expensive than the more \hbox{popular reductions.}

\section{Conclusion}
This paper presents a comprehensive overview of data reduction rules for the MWIS and MWVC problems, two fundamental $\mathsf{NP}$-hard problems with a wide range of practical applications. By presenting these reduction techniques in one place, we aim to provide researchers and practitioners with a valuable resource to simplify problem instances and enhance the performance of both exact solvers and heuristic approaches. Data reduction has proven to be a critical component for solving the MWC, MWVC, and MWIS problems, particularly in branch-and-reduce methods where effective reductions can significantly decrease the problem size and improve solvability.

As new reduction techniques continue to emerge, this work will be updated to reflect the latest advancements, ensuring that it remains a relevant and up-to-date reference for those working in this area. We will continue to develop a reference implementation for the different data reduction rules included in this survey\footnote{\url{https://github.com/KarlsruheMIS/DataReductions}}.
By centralizing this information, we aim to facilitate further progress in solving the MWIS and related problems more efficiently.

\begin{acks}
This work is supported by the \grantsponsor{funding1}{DFG}{} under grant number \grantnum{funding1}{SCHU 2567/3-1}, and \grantsponsor{funding2}{Research Council of Norway}{} under grant number \grantnum{funding2}{303404}, and \grantsponsor{funding3}{Meltzer Research Fund}{} under grant number \grantnum{funding3}{104066111}.
\end{acks}
\bibliographystyle{ACM-Reference-Format}
\bibliography{bib}

\end{document}